\documentclass[a4paper]{cas-sc}

\usepackage[numbers]{natbib}
\usepackage{graphicx}
\usepackage{amsmath,amssymb,amsfonts}
\usepackage{booktabs}
\usepackage{multirow}
\usepackage{xcolor}
\usepackage{placeins}
\usepackage{tabularx}
\usepackage{array}

\begin{document}
\let\WriteBookmarks\relax

% Float settings for the CAS single-column review layout.
\renewcommand{\topfraction}{0.90}
\renewcommand{\bottomfraction}{0.75}
\renewcommand{\textfraction}{0.08}
\renewcommand{\floatpagefraction}{0.70}
\setcounter{topnumber}{4}
\setcounter{bottomnumber}{2}
\setcounter{totalnumber}{6}

\shorttitle{Counterfactual Graph Prediction for Grid Control}
\shortauthors{F. Liu and Z. Tian}

\title[mode=title]{Counterfactual Action-Conditioned Graph Prediction for Cascading-Failure Mitigation in Power Systems}

%% Replace placeholders below with the final author information.
\author[1]{Fangxin Liu}
\cormark[1]
\ead{marialiu051@163.com}

\author[2]{Zhentao Tian}

\affiliation[1]{organization={Nanjing Institute of Technology},
    addressline={School of Traffic Engineering},
    city={Nanjing},
    state={Jiangsu},
    country={China}}

\affiliation[2]{organization={Nanjing University of Science and Technology ZiJin College},
    addressline={School of Electronic Engineering and Optoelectronic Technology},
    city={Nanjing},
    state={Jiangsu},
    country={China}}

\cortext[1]{Corresponding author}

\begin{abstract}
Cascading failures develop through coupled power flow redistribution, protection actions, and topology changes, making remedial control dependent on how alternative interventions modify the subsequent trajectory. This paper develops an action-conditioned graph predictor (ACGP) for multi-step cascade prediction and closed-loop mitigation. A same-state counterfactual branching procedure replays identical post-contingency grid states under multiple legal actions while preserving the exogenous trajectory, providing direct supervision for action-dependent electrical states, line connectivity, automatic protection trips, and terminal risk. ACGP is embedded in a receding-horizon controller that evaluates five-step candidate futures, executes the selected intervention, and replans from the observed next state. The final dataset contains 2,860 counterfactual roots and 22,880 action-conditioned branches. On the held-out test set, ACGP achieves a five-step line-loading MAE of 0.0369 and a redispatch pairwise ranking accuracy of 0.806; action ablations reduce the ranking accuracy to approximately chance level. In closed-loop Grid2Op experiments, terminal-failure and cascade rates decrease from 53.3\% and 51.1\% under do-nothing operation to 12.2\% and 7.8\%, respectively, close to the simulator-reference values of 11.1\% and 6.7\%. The results show that same-state counterfactual prediction can support effective action comparison and feedback-based cascading-failure mitigation.
\end{abstract}

% \begin{highlights}
% \item Same-state counterfactual branching isolates remedial-action effects.
% \item Action-conditioned graph prediction captures multi-step cascade evolution.
% \item Safety-oriented action ranking supports receding-horizon remedial control.
% \item Closed-loop feedback enables effective cascading-failure mitigation.
% \end{highlights}


\maketitle

% ==================================================================
\section{Introduction}
\label{sec:introduction}
% ==================================================================

The increasing penetration of distributed renewable energy (DRE) and growing exposure to extreme weather introduce additional uncertainty into transmission-system operation. The variability and forecasting errors of DRE can reduce operating margins, and weather-induced outages and protection interactions further increase exposure to cascading events~\cite{li2025study,zakariya2023systematic}. Protection behavior under DRE-rich operating conditions can also alter the propagation process~\cite{pei2026risk}. Following an initiating outage, redistributed power flows may overload neighboring lines and trigger subsequent protection actions~\cite{9272673}. The resulting sequence can develop into a large-scale disturbance, motivating continued attention to cascading-failure risk and mitigation~\cite{li2023cascading,zhao2025failure}.

Effective mitigation requires not only identifying vulnerable states but also anticipating how remedial actions alter the subsequent trajectory. Existing studies have established broad foundations for intelligent cascading-failure assessment~\cite{hayat2025expert} and quantitative security evaluation~\cite{10497182}. Vulnerable sequence identification provides additional information on likely propagation paths~\cite{10931178}, while preventive control studies demonstrate that corrective actions can substantially change cascade outcomes~\cite{deng2025cascading}. Prediction and control, however, are still often developed as separate functions, leaving limited capability for rapidly comparing multiple action dependent futures from the same stressed operating state. Physics-based and probabilistic cascade models remain central to understanding propagation mechanisms. Steady-state formulations have been used to represent sequential component outages~\cite{10294236}, while influence-based models characterize dependencies among outage events~\cite{9384281}. Probabilistic models further describe cascade evolution under changing operating conditions~\cite{9050711}, and discrete-event formulations provide a structured representation of sequential failure and mitigation processes~\cite{9733408}. Such approaches provide interpretable cascade analysis, but they are generally not designed as reusable learned transition models for repeated comparison of several remedial actions from an identical stressed state.

Data-driven methods provide faster alternatives for cascade screening and event detection~\cite{10849943}. Deep event-detection models have also been developed for online cascading event assessment~\cite{tang2025deep}. Sequential pattern mining supports the identification of recurring propagation structures~\cite{9215038}. Graph-based learning is particularly relevant because transmission networks possess an explicit electrical and topological structure. Physics-informed GNNs exploit this structure for power system prediction~\cite{9881910}, while geometric deep learning has been applied to online cascade forecasting~\cite{varbella2023geometric}. Graph models have also been used for line status prediction~\cite{11653266}, critical component identification~\cite{soleimani2025critical}, anomalous node localization~\cite{zhao2026anomalous}, and critical node assessment~\cite{zhang2026egds}. Most of these prediction-oriented methods, however, do not explicitly condition multi-step future evolution on alternative remedial actions.

A complementary research stream addresses cascade mitigation through economic dispatch~\cite{deng2025economic} and sequential preventive optimization~\cite{liu2022sequentially}. Coordination control can block propagation after line outages~\cite{10374403}, while adaptive network response strategies exploit corrective network actions~\cite{9872085}. Data-driven surrogates have also begun to support security-oriented decision problems~\cite{11654520}. These methods demonstrate the value of corrective control, but the selected action is commonly obtained through repeated simulation, problem specific optimization, or a task specific surrogate rather than through multi-step prediction of several alternative futures from an identical operating state. The reviewed studies therefore establish strong foundations for cascade simulation, graph-based prediction, and remedial control, but these capabilities are still commonly developed separately. Online mitigation requires action-conditioned multi-step prediction and feedback-based replanning from the observed grid state. Table~\ref{tab:knowledgegap} summarizes this methodological gap.

\begin{table}[pos=htbp]
\caption{Comparison of representative cascading-failure prediction and control methods.}
\label{tab:knowledgegap}
\centering
\scriptsize
\resizebox{\linewidth}{!}{%
\begin{tabular}{lccccccc}
\toprule
Reference
& \shortstack{Cascade\\analysis}
& \shortstack{Graph\\learning}
& \shortstack{Sequential\\modeling}
& \shortstack{Remedial\\control}
& \shortstack{Action-conditioned\\prediction}
& \shortstack{Same-state\\branching}
& \shortstack{Closed-loop\\replanning} \\
\midrule
Varbella et al.~\cite{varbella2023geometric}
& $\checkmark$ & $\checkmark$ & $\times$ & $\times$ & $\times$ & $\times$ & $\times$ \\
Zhu et al.~\cite{9881910}
& $\checkmark$ & $\checkmark$ & $\checkmark$ & $\times$ & $\times$ & $\times$ & $\times$ \\
Sami and Naeini~\cite{11653266}
& $\checkmark$ & $\checkmark$ & $\times$ & $\times$ & $\times$ & $\times$ & $\times$ \\
He et al.~\cite{10931178}
& $\checkmark$ & $\times$ & $\checkmark$ & $\times$ & $\times$ & $\times$ & $\times$ \\
Deng et al.~\cite{deng2025cascading}
& $\checkmark$ & $\times$ &$\times$ & $\checkmark$ & $\times$ & $\times$ & $\times$ \\
Ouyang et al.~\cite{10374403}
& $\checkmark$ & $\times$ & $\times$ & $\checkmark$ & $\times$ & $\times$ & $\times$ \\
Hu et al.~\cite{11654520}
& $\checkmark$ & $\times$ & $\times$ & $\checkmark$ & $\times$ & $\times$ & $\times$ \\
Al-Rousan et al.~\cite{9733408}
& $\checkmark$ & $\times$ & $\checkmark$ & $\checkmark$ & $\times$ & $\times$ & $\checkmark$ \\
\midrule
\textbf{This work (ACGP)}
& $\checkmark$ & $\checkmark$ & $\checkmark$ & $\checkmark$ & $\checkmark$ & $\checkmark$ & $\checkmark$ \\
\bottomrule
\end{tabular}}
\end{table}

To address this gap, we develop an \emph{Action-Conditioned Graph Predictor (ACGP)} for multi-step prediction of controlled cascading-failure trajectories. Given the current graph state and a legal remedial action, ACGP predicts future electrical states, line connectivity, automatic protection trips, and terminal risk. Same-state counterfactual branches are constructed by replaying an identical post-contingency root while preserving the chronic, initiating contingency, and pre-root trajectory, thereby providing direct supervision for the effect of alternative interventions. The trained ACGP is embedded in a receding-horizon controller. At each post-contingency decision step, legal do-nothing, redispatch, and reconnection candidates are evaluated through five-step predictive rollouts and ranked by terminal-failure risk, cascade risk, and maximum line loading. Only the selected action is executed in the Grid2Op simulation environment~\cite{grid2op}; the resulting physical state is then observed and used for the next receding-horizon decision. The main contributions are summarized as follows:
\begin{enumerate}
\item Cascading-failure evolution is formulated as an action-conditioned multi-step graph prediction problem that jointly represents continuous electrical states, line connectivity, automatic protection trips, and terminal risk.

\item A same-state counterfactual branching procedure is developed to isolate the effects of legal redispatch and reconnection actions under identical exogenous conditions, together with neutral-aware redispatch-effect learning for fine-grained same-root action ranking.

\item ACGP is integrated into a safety-first receding-horizon control framework that predicts candidate futures, executes only the selected intervention, and replans from the subsequently observed grid state.

\item The complete prediction-to-control pipeline is evaluated through held-out multi-step prediction, action-input ablations, same-root ranking tests, and closed-loop cascading-failure experiments, demonstrating substantial reductions in terminal failure and cascade occurrence relative to do-nothing operation.
\end{enumerate}

% ==================================================================
\section{Problem Formulation and Control Objective}
\label{sec:problem}
% ==================================================================

\begin{figure}[pos=htbp]
\centering
\includegraphics[width=\linewidth]{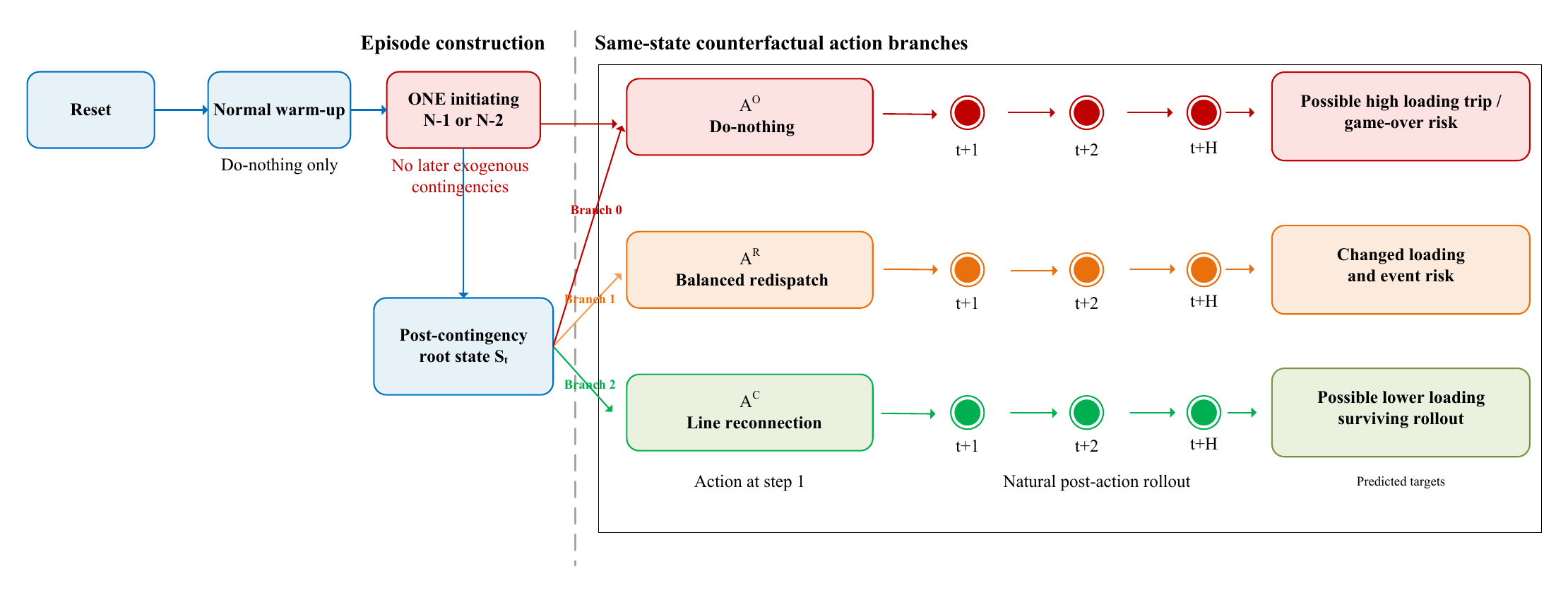}
\vspace{-0.7cm}
\caption{Problem setting and same-state counterfactual branching. Following one initiating N-1 or N-2 contingency, the same post-contingency state $S_t$ is branched under do-nothing, balanced-redispatch, and legal-reconnection actions. Subsequent steps follow the canonical do-nothing rollout.}
\vspace{-0.5cm}
\label{fig:problem-setting}
\end{figure}

Figure~\ref{fig:problem-setting} summarizes the data-generating and decision setting. Each episode contains one initiating contingency. Counterfactual branches start from the same observed post-contingency root and differ only in their first legal remedial action; subsequent rollout steps use the canonical do-nothing action.

\subsection{Cascading-failure setting and counterfactual branches}

At decision time $t$, the transmission grid is represented by the graph defined in \eqref{eq1}, where substations constitute the node set $\mathcal{V}$ and physical transmission lines constitute the edge set $\mathcal{E}$. The matrices $\mathbf{X}^{v}_t$ and $\mathbf{X}^{e}_t$ collect the observable node and edge features, respectively.
\begin{equation}
\label{eq1}
    \mathcal{G}_t=(\mathcal{V},\mathcal{E},\mathbf{X}^{v}_t,\mathbf{X}^{e}_t),
\end{equation}
The physical graph structure remains fixed, with disconnected lines retained as edges and their connectivity represented by line-status and operating features. The controller observation is:
\begin{equation}
    S_t=(\mathcal{G}_t,\mathbf{z}_t),
\end{equation}
where $\mathbf{z}_t$ contains the topology vector and simulator observables not represented directly by the graph features.

Each episode begins with a reset and an optional warm-up period, followed by a single initiating N-1 or N-2 outage and the subsequent natural cascade evolution. The initiating outage is treated as an exogenous disturbance rather than a control action. No additional exogenous outages are introduced after this event. Any later line disconnections are triggered by the simulator protection logic and are recorded separately as automatic trips. For a fixed chronic and initiating contingency $C$, the physical transition is written as
\begin{equation}
    (S_{t+1},\mathbf{m}^{\mathrm{trip}}_{t+1},d_{t+1})
    =F(S_t,A_t,C,\xi_t),
    \label{eq:physical-transition}
\end{equation}
where $A_t$ is a remedial action, $\xi_t$ denotes exogenous chronic evolution, $\mathbf{m}^{\mathrm{trip}}_{t+1}$ is the automatic-trip mask, and $d_{t+1}$ indicates terminal failure. Branches terminating before horizon $H$ are masked beyond their last valid transition.

At a post-contingency root state $S_t$, the controller has a finite set of legal actions $\mathcal{A}(S_t)$ containing do nothing, balanced redispatch, and legal line reconnection. For candidate $A_t^{(k)}$, the counterfactual sequence applies the intervention once and then lets the grid evolve:
\begin{equation}
    \mathbf{A}^{(k)}_{t:t+H-1}
    =[A_t^{(k)},A^0,\ldots,A^0],
    \label{eq:branch-action-sequence}
\end{equation}
where $A^0$ is the canonical do-nothing action.

All counterfactual branches preserve the same root state, chronic, initiating contingency, and exogenous trajectory. The corresponding invariance condition and horizon-level maximum loading are defined as \eqref{eq5} and \eqref{eq6}.
\begin{align}
    (S_t^{(k)},C^{(k)},\xi_{0:t+H}^{(k)})
    &=(S_t,C,\xi_{0:t+H}), \qquad \forall k,
    \label{eq5}
\end{align}
\vspace{-0.3cm}
\begin{align}    
    \rho^{(k)}_{\max,H}
    &=\max_{\tau=1,\ldots,H}\max_{\ell\in\mathcal{E}}
    \rho^{(k)}_{t+\tau,\ell}.
    \label{eq6}
\end{align}

\subsection{Action-conditioned prediction and receding-horizon control}

The proposed ACGP $f_\theta$ predicts the controlled multi-step future, including electrical quantities, line status, automatic trips, and terminal risk. Predicted states are recursively fed back during rollout, so the model must remain stable beyond one-step prediction.
\begin{equation}
    \widehat{\mathbf{Y}}^{(k)}_{t+1:t+H}
    =f_\theta(S_t,\mathbf{A}^{(k)}_{t:t+H-1}),
    \qquad H=5,
    \label{eq:predictor-rollout}
\end{equation}

The planner ranks the candidate futures using the lexicographic safety tuple defined in \eqref{eq8}.
\begin{equation}
    J^{(k)}=
    \left(
    \hat g^{(k)},
    \hat c^{(k)},
    \max_{\tau,\ell}\hat{\rho}^{(k)}_{t+\tau,\ell}
    \right),
    \label{eq8}
\end{equation}
where $\hat g^{(k)}$ and $\hat c^{(k)}$ denote the predicted terminal-failure and cascade classes, respectively. Their decision thresholds are determined using the validation set. Predicted trip count is retained as a diagnostic quantity but is excluded from the final safety criterion because its same-root ranking was not sufficiently reliable during validation.

The candidate selected for execution is determined according to \eqref{eq9}.
\begin{equation}
    k_t^*=
    \underset{k:\,A_t^{(k)}\in\mathcal{A}(S_t)}
    {\arg\min}\; J^{(k)},
    \label{eq9}
\end{equation}

The tuple $J^{(k)}$ is compared lexicographically, with terminal-failure class assigned the highest priority, followed by cascade class and maximum line loading. The corresponding remedial action is $A_t^*=A_t^{(k_t^*)}$. After its execution, the resulting grid state is observed and used for the next receding-horizon decision.

% ==================================================================
\section{Action-Conditioned Graph Prediction Framework}
\label{sec:method}
% ==================================================================

\begin{figure}[pos=htbp]
\centering
\includegraphics[width=0.7\linewidth]{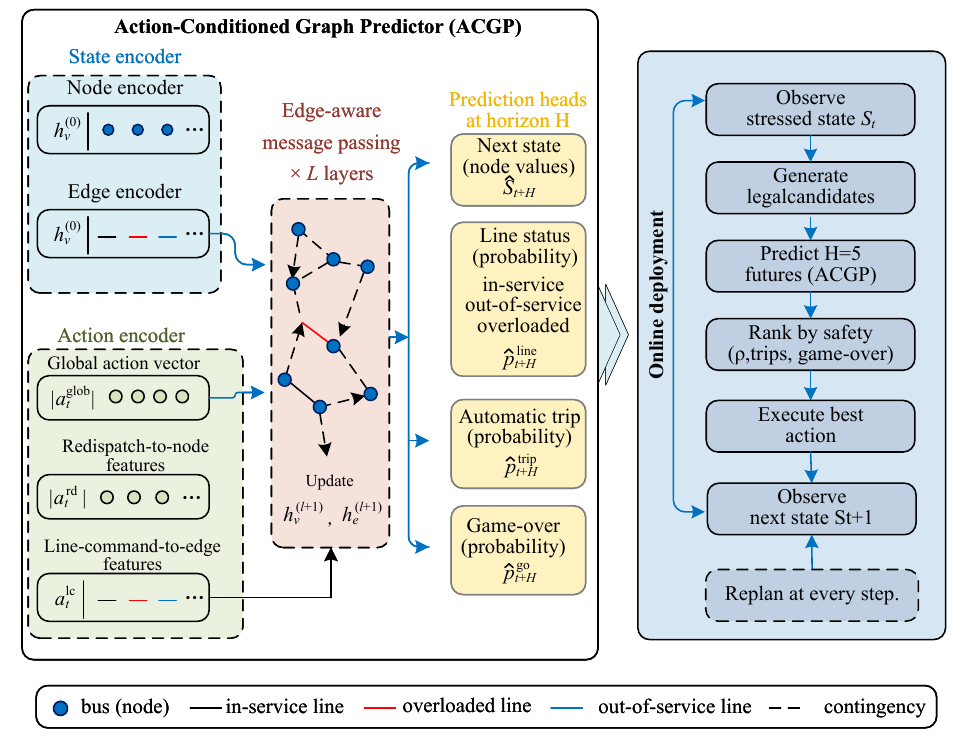}
\caption{Overview of ACGP and its closed-loop use. Legal candidate actions define alternative futures from the same post-contingency state; ACGP predicts their multi-step evolution, and the controller executes the safest candidate before replanning.}
\label{fig:acgp-workflow}
\vspace{-0.5cm}
\end{figure}

Figure~\ref{fig:acgp-workflow} provides an overview of the proposed framework. Grid2Op observations are converted into graph-structured states, and same-state replay generates legal action-conditioned branches. ACGP then predicts the multi-step evolution of electrical states, network connectivity, protection events, and terminal risk. The predicted trajectories are used to rank the candidate actions, after which only the selected intervention is executed in the simulator before replanning.

\subsection{State-action representation and counterfactual branch construction}

Each Grid2Op observation is mapped to a graph-structured representation in which substations correspond to nodes and physical transmission lines correspond to edges. Seven node features describe the aggregated load and generation conditions, generator voltage, and connection information, while thirteen edge features characterize terminal power flows, electrical quantities, loading conditions, line status, and protection-related states. The complete feature set is summarized in Table~\ref{tab:graph-features}.

\begin{table}[pos=htbp]
\caption{Node and edge features used by ACGP.}
\label{tab:graph-features}
\centering
\small
\begin{tabularx}{\linewidth}{
@{}>{\centering\arraybackslash}p{0.10\linewidth}
>{\raggedright\arraybackslash}X
>{\centering\arraybackslash}p{0.13\linewidth}
>{\raggedright\arraybackslash}X@{}
}
\toprule
\multicolumn{2}{c}{Node features}
& \multicolumn{2}{c}{Edge features}\\
\cmidrule(r){1-2}
\cmidrule(l){3-4}
Index & Feature & Index & Feature\\
\midrule
1 & Aggregated active load
& 1--4 & $P,Q,V,I$ at line origin\\

2 & Aggregated reactive load
& 5--8 & $P,Q,V,I$ at line extremity\\

3 & Aggregated active generation
& 9 & Loading ratio $\rho$\\

4 & Aggregated reactive generation
& 10 & Thermal limit\\

5 & Mean generator voltage
& 11 & Line status\\

6 & Number of connected loads
& 12 & Overflow duration\\

7 & Number of connected generators
& 13 & Line cooldown\\
\bottomrule
\end{tabularx}
\end{table}

To avoid information leakage, all continuous graph features are standardized using statistics computed exclusively from the training split. The normalized node and edge features are expressed as
\begin{align}
    \widetilde{\mathbf{x}}^{v}_{t,i}
    &=\frac{\mathbf{x}^{v}_{t,i}-\boldsymbol{\mu}_{v}}
    {\boldsymbol{\sigma}_{v}+\varepsilon},\\
    \widetilde{\mathbf{x}}^{e}_{t,\ell}
    &=\frac{\mathbf{x}^{e}_{t,\ell}-\boldsymbol{\mu}_{e}}
    {\boldsymbol{\sigma}_{e}+\varepsilon}.
    \label{eq:feature-normalization}
\end{align}

Each action is represented at three levels. The full Grid2Op action vector and action-family identifier form a global action feature; requested redispatch is projected from generators to substations; and line-status commands are represented as edge-action features. No future-derived field, including cascade labels, time-to-failure, or branch-improvement labels, is used as model input.

For action vector $\mathbf{a}_t$, action family $q_t$, redispatch command $\Delta P_{t,g}$, and generator-to-substation mapping $\pi(g)$,
\begin{align}
    \mathbf{u}^{A}_t &= \phi_{\mathrm{global}}([\mathbf{a}_t;\mathrm{onehot}(q_t)]), \\
    r^{A}_{t,i} &= \sum_{g:\pi(g)=i}\Delta P_{t,g}, \\
    e^{A}_{t,\ell} &= a_{t,\ell}-a^{0}_{\ell},
    \label{eq:action-projection}
\end{align}
where $\mathbf{u}^{A}_t$, $r^{A}_{t,i}$, and $e^{A}_{t,\ell}$ are the global, node, and edge action features. The initial encodings are
\begin{align}
    \mathbf{h}^{(0)}_{t,i}
    &=\phi_v([\mathbf{x}^{v}_{t,i};r^A_{t,i};\mathbf{u}^A_t]),\\
    \mathbf{z}^{(0)}_{t,\ell}
    &=\phi_e([\mathbf{x}^{e}_{t,\ell};e^A_{t,\ell};\mathbf{u}^A_t]).
    \label{eq:action-conditioned-encodings}
\end{align}

Training samples are generated from post-contingency trajectories obtained in Grid2Op with the LightSim2Grid backend. Each trajectory contains a single initiating contingency followed by natural system evolution, with initiating outages distinguished from subsequent protection-triggered line trips. Valid post-contingency states are then used as roots for same-state counterfactual branching. Each root is reproduced under identical exogenous conditions, and multiple legal remedial actions are evaluated from independently replayed copies of the same state. After the initial intervention, the system evolves under do-nothing operation over the remaining prediction horizon. This construction provides action-conditioned trajectories in which differences among branches can be attributed to the applied remedial action rather than to changes in the initial state or exogenous evolution.

\subsection{Action-Conditioned Graph Predictor}

ACGP predicts the subsequent grid evolution conditioned on the current operating state and the applied remedial action. State and action representations are jointly embedded into an edge-aware graph architecture to capture their influence on power-system dynamics. At graph layer $r$, information associated with physical line $\ell=(i,j)$ is propagated between connected substations according to equations \eqref{eq17}--\eqref{eq19}.
\begin{align}
    \mathbf{m}^{(r)}_{t,\ell}
    &=\psi_r\!\left(
    \mathbf{h}^{(r)}_{t,i},\mathbf{h}^{(r)}_{t,j},
    \mathbf{z}^{(r)}_{t,\ell},\mathbf{u}^{A}_t\right),
    \label{eq17}\\
    \mathbf{h}^{(r+1)}_{t,i}
    &=\eta_r\!\left(
    \mathbf{h}^{(r)}_{t,i},
    \sum_{\ell\in\mathcal{N}(i)}\mathbf{m}^{(r)}_{t,\ell}
    \right),
    \label{eq18}\\
    \mathbf{z}^{(r+1)}_{t,\ell}
    &=\zeta_r\!\left(
    \mathbf{z}^{(r)}_{t,\ell},
    \mathbf{h}^{(r+1)}_{t,i},
    \mathbf{h}^{(r+1)}_{t,j}\right).
    \label{eq19}
\end{align}

The shared graph representation feeds continuous node/edge decoders, a line-status head, an automatic-trip head, and a graph-level terminal head. Their one-step outputs are written as equations \eqref{eq20}-\eqref{eq23} and $R$ denotes the final message-passing layer.
\begin{align}
    \widehat{\mathbf{X}}^{v}_{t+1}
    &=D_v\!\left(\mathbf{H}^{(R)}_t\right),
    \label{eq20}\\
    \widehat{\mathbf{X}}^{e}_{t+1}
    &=D_e\!\left(\mathbf{Z}^{(R)}_t\right),
    \label{eq21}\\
    \left(\widehat{\mathbf{p}}^{\,\mathrm{status}}_{t+1},
    \widehat{\mathbf{p}}^{\,\mathrm{trip}}_{t+1}\right)
    &=D_{\mathrm{edge}}\!\left(\mathbf{Z}^{(R)}_t\right),
    \label{eq22}\\
    \widehat p^{\,\mathrm{term}}_{t+1}
    &=D_{\mathrm{term}}\!\left(\mathbf{H}^{(R)}_t,\mathbf{Z}^{(R)}_t\right).
    \label{eq23}
\end{align}

For recursive rollout, the predicted quantities are assembled into the next model state before the following action is evaluated via \eqref{eq24}. $\mathcal{R}(\cdot)$ denotes deterministic reconstruction of the graph state used as the next model input.
\begin{equation}
    \widehat S_{t+\tau}
    =\mathcal{R}\!\left(
    \widehat{\mathbf{X}}^{v}_{t+\tau},
    \widehat{\mathbf{X}}^{e}_{t+\tau},
    \widehat{\mathbf{p}}^{\,\mathrm{status}}_{t+\tau}
    \right),
    \qquad \tau=1,\ldots,H ,
    \label{eq24}
\end{equation}

Training minimizes the overall objective $\mathcal{L}$, which combines the one-step anchor loss $\mathcal{L}_{1}$ and the closed-loop rollout loss $\mathcal{L}_{\mathrm{roll}}$.
\begin{equation}
    \mathcal{L}
    =\mathcal{L}_{1}+0.25\,\mathcal{L}_{\mathrm{roll}} .
    \label{eq:training-loss}
\end{equation}

At each prediction step, the objective $\ell_{t+\tau}$ combines the node-state loss $\mathcal{L}_{v}$, edge-state loss $\mathcal{L}_{e}$, line-loading loss $\mathcal{L}_{\rho}$, line-status loss $\mathcal{L}_{\mathrm{status}}$, protection-trip loss $\mathcal{L}_{\mathrm{trip}}$, and terminal-failure loss $\mathcal{L}_{\mathrm{term}}$. The line-loading term $\mathcal{L}_{\rho}$ is assigned a weight of two to place greater emphasis on loading prediction.
\begin{equation}
    \ell_{t+\tau}
    =
    \lambda_v\mathcal{L}_{v}
    +\lambda_e\mathcal{L}_{e}
    +2\mathcal{L}_{\rho}
    +\lambda_s\mathcal{L}_{\mathrm{status}}
    +\lambda_{\mathrm{tr}}\mathcal{L}_{\mathrm{trip}}
    +\lambda_g\mathcal{L}_{\mathrm{term}} .
    \label{eq:per-step-objective}
\end{equation}

Continuous outputs use regression losses; discrete outputs use weighted binary cross-entropy with weights estimated from the training split. With future-valid mask $v_{t+\tau}$,
\begin{equation}
    \mathcal{L}_{\mathrm{roll}}
    =
    \frac{\sum_{\tau=1}^{H}0.9^{\tau-1}v_{t+\tau}\ell_{t+\tau}}
    {\sum_{\tau=1}^{H}0.9^{\tau-1}v_{t+\tau}} .
    \label{eq:masked-rollout-loss}
\end{equation}

Predicted states are recursively fed back without teacher forcing after the root state, and a curriculum increases the rollout horizon from one to five steps.
\subsection{Neutral-aware redispatch effect modeling}

Redispatch actions may induce only marginal changes in system loading, making their relative effects difficult to distinguish from near-neutral responses. To characterize the action effect under an identical operating condition, the change in maximum loading is measured with respect to the corresponding do-nothing branch as \eqref{eq28}.The three-class redispatch label is \eqref{eq29}
\begin{equation}
    \Delta\rho_{\max}
    =\rho_{\max}^{\mathrm{redispatch}}
    -\rho_{\max}^{\mathrm{noop}} .
    \label{eq28}
\end{equation}
\begin{equation}
    y_{\Delta\rho}=
    \begin{cases}
    \mathrm{improve}, & \Delta\rho_{\max}<-10^{-4},\\
    \mathrm{neutral}, & |\Delta\rho_{\max}|\le10^{-4},\\
    \mathrm{worsen}, & \Delta\rho_{\max}>10^{-4}.
    \end{cases}
    \label{eq29}
\end{equation}

Changes within the neutral tolerance band are excluded from pairwise ordering, avoiding artificial preference assignments between nearly equivalent redispatch outcomes. For two non-tied redispatch candidates $i$ and $j$, the pairwise target is defined as
\begin{equation}
    y_{ij}
    =\mathbb{I}\!\left[
    \Delta\rho_{\max}^{(i)}
    <\Delta\rho_{\max}^{(j)}
    \right],
    \label{eq:pairwise-target}
\end{equation}
where $y_{ij}=1$ indicates that candidate $i$ yields a smaller increase, or a larger reduction, in maximum line loading than candidate $j$.

An auxiliary redispatch-effect module jointly predicts the three-class action effect, the magnitude of $\Delta\rho_{\max}$, and the relative preference between non-tied candidates. The pairwise ranking loss is defined as
\begin{equation}
    \mathcal{L}_{\mathrm{rank}}
    =-\frac{1}{|\mathcal{P}|}
    \sum_{(i,j)\in\mathcal{P}}
    \left[
    y_{ij}\log \sigma(\hat r_j-\hat r_i)
    +(1-y_{ij})\log \sigma(\hat r_i-\hat r_j)
    \right],
    \label{eq31}
\end{equation}
where $\mathcal{P}$ denotes the set of non-tied redispatch pairs, $\sigma(\cdot)$ is the sigmoid function, and $\hat r$ is the predicted ranking score. A lower value of $\hat r$ corresponds to a more favorable predicted loading effect. During training of the redispatch-effect module, the graph-dynamics backbone remains fixed.

% \subsection{Receding-horizon planner}

At each post-contingency control step, the planner observes the current grid state, constructs the legal candidate set, and predicts a five-step trajectory for each candidate. The candidates are ranked using the lexicographic safety tuple in Eq.~\eqref{eq8}, and the control action is selected according to Eq.~\eqref{eq9}. Terminal-failure and cascade classes are determined using event thresholds calibrated on the validation set. Only the selected action is executed in the simulator, after which the resulting grid state is observed and used for the next receding-horizon decision. Future simulator trajectories are not available to ACGP during action selection. For comparison, a separate simulator-reference planner evaluates the same candidate set over the same prediction horizon using true simulator rollouts and is used as an evaluation benchmark.

% ==================================================================
\section{Experimental Results and Discussion}
\label{sec:setup}
% ==================================================================

\subsection{Experimental setup and training diagnostics}

Experiments are conducted in the Grid2Op \texttt{l2rpn\_case14\_sandbox} environment using the LightSim2Grid backend. The system contains 14 substations, 20 transmission lines, 11 loads, and 6 generators. The counterfactual dataset comprises 2,860 roots and 22,880 action-conditioned branches, with 1,996/422/442 roots assigned to the training, validation, and test sets by source episode. No source episode, chronic trajectory, or root is shared across splits. Normalization statistics and class weights are obtained from the training set, while decision thresholds are calibrated on the validation set.

Data consistency is verified through exact root replay, factual do-nothing reproduction over the five-step horizon, and branch-order invariance, with no failed checks observed. ACGP is initialized from the stabilized factual predictor and trained with a learning rate of $10^{-4}$ and a batch size of 512. The graph-dynamics backbone contains 243,179 parameters and the neutral-aware redispatch-effect module contains 36,036 parameters, with their checkpoints selected at epochs 23 and 15, respectively.

Closed-loop evaluation compares do-nothing control, ACGP, and an evaluation-only simulator-reference planner using the same initial states and legal candidate sets. ACGP and the simulator reference evaluate candidate actions over a five-step horizon using the safety criterion in Eq.~\eqref{eq8}. Only the selected action is executed before replanning from the next observed grid state. Five of the 90 held-out episodes terminate during the initiating contingency and are retained as common uncontrollable outcomes. Grid2Op's ``game-over'' condition is reported as \emph{terminal failure}.

Prediction performance is evaluated using line-loading MAE at horizons one, three, and five, same-root pairwise ranking accuracy, rank correlation, effect-magnitude error, Top-1/Top-3 recovery, and safety regret. Redispatch pairs satisfying $|\Delta\rho_{\max}|\le10^{-4}$ are treated as ties. Closed-loop evaluation considers terminal-failure and cascade rates, automatic trips, survival duration, maximum line loading, intervention effort, rescue and cascade-prevention rates, and harmful interventions. Action ablations replace the true action inputs with no-op or shuffled alternatives.

\begin{figure}[pos=ht!]
\centering
\begin{minipage}[t]{0.48\linewidth}
\centering
\includegraphics[width=0.85\linewidth]{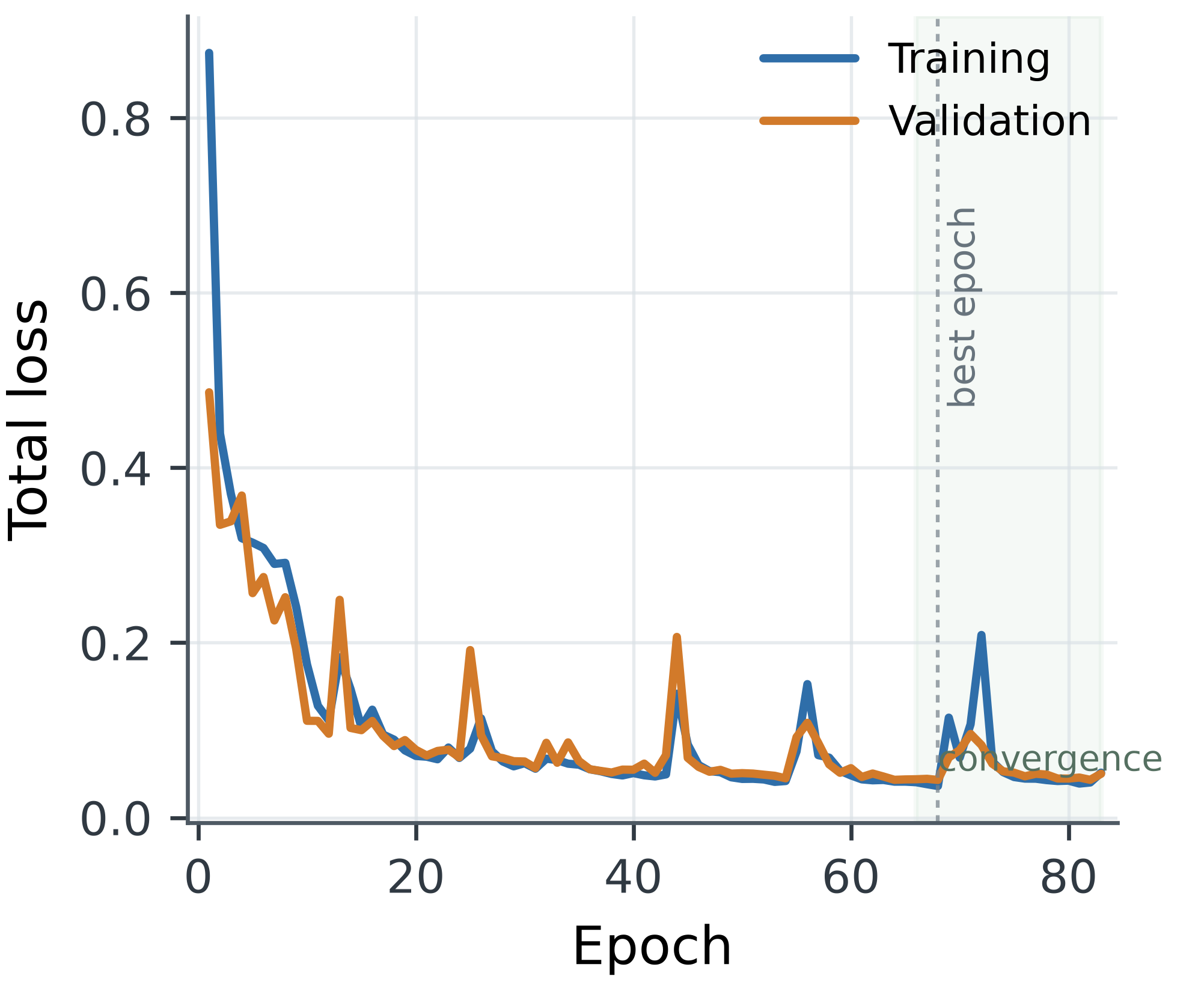}\\[-0.3em]
\textbf{(a)}
\end{minipage}
\hfill
\begin{minipage}[t]{0.48\linewidth}
\centering
\includegraphics[width=0.85\linewidth]{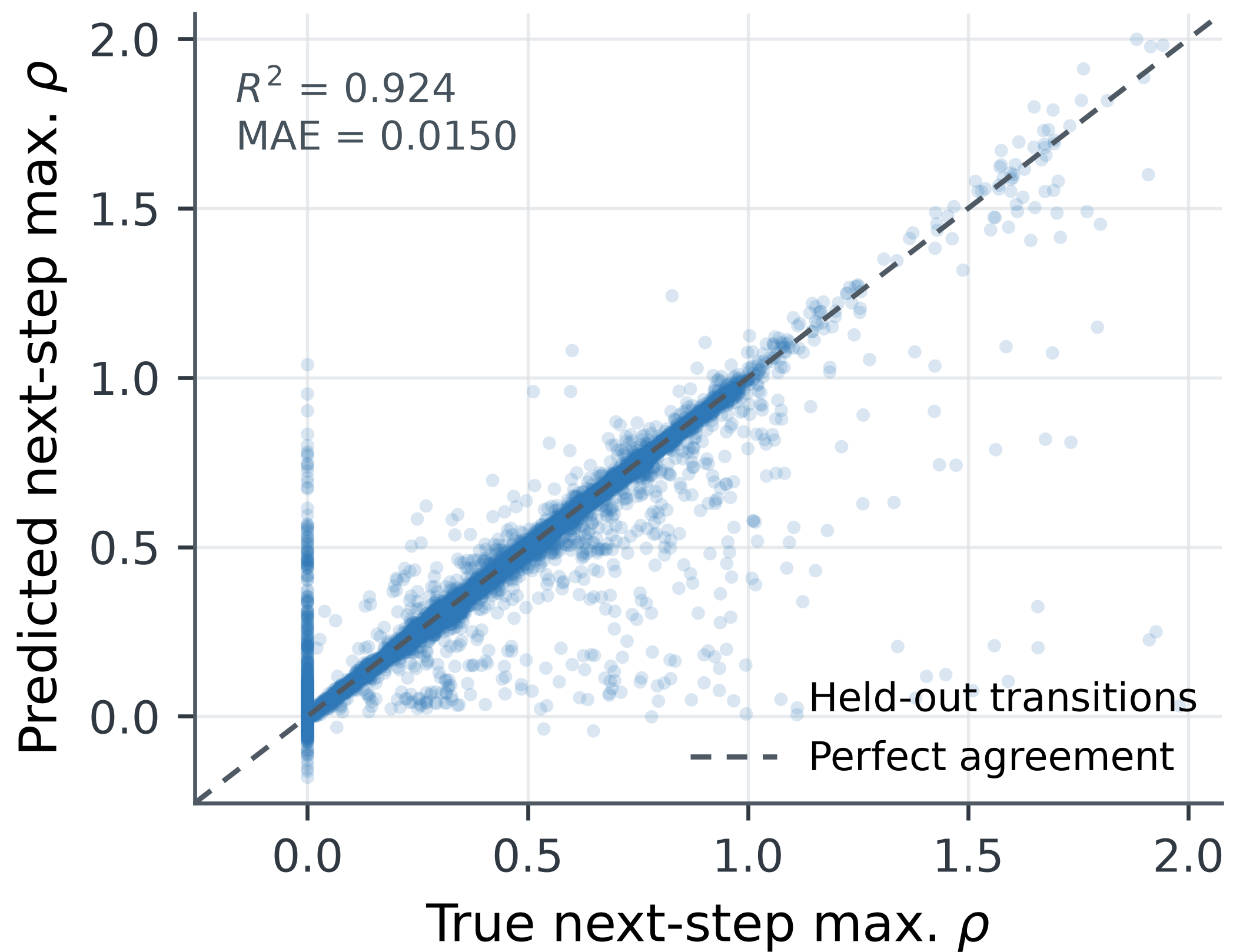}\\[-0.3em]
\textbf{(b)}
\end{minipage}
\caption{Training and held-out prediction diagnostics: (a) training and validation loss; (b) simulated versus predicted next-step maximum line loading.}
\label{fig:trainingdiagnostics}
\end{figure}

Figure~\ref{fig:trainingdiagnostics} shows the training behavior and held-out next-step loading prediction used to verify stable optimization before the multi-step evaluation. The training and validation curves decrease consistently, and the held-out prediction cloud follows the simulated loading trend. These diagnostics support the numerical stability of the learned transition model; the more decision-relevant multi-step and ranking results are examined next.
\vspace{-0.5cm}
\FloatBarrier

\subsection{Multi-step prediction and action ranking}

Table~\ref{tab:prediction} summarizes the held-out multi-step prediction and action-ranking performance. The line-loading MAE increases from 0.02998 at $H=1$ to 0.03687 at $H=5$, indicating limited error accumulation over the five-step rollout. For the 2,569 non-tied redispatch pairs, ACGP achieves a pairwise ranking accuracy of 0.8060, together with positive Spearman and Kendall rank correlations. The all-action pairwise accuracy of 0.7169 and Top-3 recovery rate of 0.8801 further show that the predicted trajectories preserve sufficient relative ordering among candidate actions for control-oriented comparison, even when the highest-ranked action is not always recovered.

\begin{table}[pos=ht!]
\caption{Held-out multi-step prediction and action-ranking performance.}
\label{tab:prediction}
\centering
\small
\begin{tabularx}{\linewidth}{
@{}>{\raggedright\arraybackslash}X
>{\centering\arraybackslash}p{0.14\linewidth}
>{\raggedright\arraybackslash}X
>{\centering\arraybackslash}p{0.14\linewidth}@{}
}
\toprule
\multicolumn{2}{c}{Prediction and effect accuracy}
& \multicolumn{2}{c}{Action ranking and safety}\\
\cmidrule(r){1-2}
\cmidrule(l){3-4}
Metric & Result & Metric & Result\\
\midrule
$H=1$ $\rho$ MAE & 0.02998
& Redispatch non-tie pairs & 2,569\\

$H=3$ $\rho$ MAE & 0.03603
& Redispatch pairwise accuracy & 0.8060\\

$H=5$ $\rho$ MAE & 0.03687
& Spearman correlation & 0.5834\\

$\Delta\rho_{\max}$ MAE & 0.01670
& Kendall correlation & 0.5175\\

Terminal-failure regret & 0.0339
& All-action pairwise accuracy & 0.7169\\

Cascade regret & 0.0339
& Top-1 / Top-3 recovery & 0.6810 / 0.8801\\
\bottomrule
\end{tabularx}
\end{table}

\FloatBarrier

Table~\ref{tab:action-ablation} examines whether the observed ranking performance is genuinely driven by the remedial-action input. Replacing all test actions with do nothing reduces both pairwise metrics to 0.5000, while shuffling the action representations produces a comparable degradation. These results show that the same-root ordering cannot be inferred from the stressed grid state alone and instead depends on the encoded intervention. This distinction is essential for control, where the planner must discriminate among alternative remedial actions applied to the same operating condition rather than merely estimate overall system severity. Predicted trip count is retained only as a diagnostic because its inclusion degraded same-root action selection during validation. The final safety criterion therefore consists of terminal-failure class, cascade class, and maximum line loading.
\vspace{-0.3cm}
\begin{table}[pos=ht!]
\caption{Action-input ablation on the held-out counterfactual test split.}
\label{tab:action-ablation}
\centering
\small
\begin{tabularx}{\linewidth}{
@{}>{\raggedright\arraybackslash}X
>{\centering\arraybackslash}p{0.23\linewidth}
>{\centering\arraybackslash}p{0.23\linewidth}
>{\centering\arraybackslash}p{0.20\linewidth}@{}
}
\toprule
Input mode & Redispatch pairwise & All-action pairwise & Top-1 recovery\\
\midrule
True action & \textbf{0.8060} & \textbf{0.7169} & \textbf{0.6810}\\
Action zeroed & 0.5000 & 0.5000 & 0.0317\\
Action shuffled & 0.4889 & 0.4688 & 0.1674\\
\bottomrule
\end{tabularx}
\end{table}

\FloatBarrier

\subsection{Closed-loop mitigation performance}

Table~\ref{tab:control} summarizes the main closed-loop results. Compared with do-nothing operation, ACGP reduces the terminal-failure rate from 53.3\% to 12.2\% and the cascade rate from 51.1\% to 7.8\%, close to the simulator-reference values of 11.1\% and 6.7\%. The mean number of automatic trips decreases from 1.211 to 0.122, while mean survival increases from 5.84 to 8.91, confirming that the improvement extends beyond terminal outcomes. By contrast, the changes in loading statistics are more moderate, indicating that effective mitigation depends not only on reducing overall loading levels but also on preventing critical transitions that would otherwise sustain cascade propagation.

\begin{table}[pos=ht!]
\caption{Closed-loop control performance on held-out episodes.}
\label{tab:control}
\centering
\small
\begin{tabularx}{\linewidth}{
@{}>{\raggedright\arraybackslash}X
>{\centering\arraybackslash}p{0.18\linewidth}
>{\centering\arraybackslash}p{0.18\linewidth}
>{\centering\arraybackslash}p{0.22\linewidth}@{}
}
\toprule
Metric & Do nothing & ACGP & Simulator ref.\\
\midrule
Terminal-failure rate & 53.3\% & \textbf{12.2\%} & 11.1\%\\
Cascade rate & 51.1\% & \textbf{7.8\%} & 6.7\%\\
Mean automatic trips & 1.211 & \textbf{0.122} & 0.133\\
Mean survival steps & 5.84 & \textbf{8.91} & 9.00\\
Mean max. $\rho$ & 1.207 & \textbf{1.167} & 1.159\\
P95 max. $\rho$ & 1.794 & \textbf{1.738} & 1.708\\
Non-no-op actions/episode & 0.00 & \textbf{4.48} & 8.94\\
\bottomrule
\end{tabularx}
\end{table}

\FloatBarrier

\begin{table}[pos=ht!]
\caption{Safety intervention and action selectivity on held-out source episodes.}
\label{tab:rescue-selectivity}
\centering
\small
\begin{tabularx}{\linewidth}{
@{}>{\raggedright\arraybackslash}X
>{\centering\arraybackslash}p{0.24\linewidth}
>{\centering\arraybackslash}p{0.24\linewidth}@{}
}
\toprule
Metric & ACGP & Simulator ref.\\
\midrule
Blackout rescue & 37/48 (77.1\%) & 38/48 (79.2\%)\\
Cascade prevention & 39/46 (84.8\%) & 40/46 (87.0\%)\\
Harmful control & 0/42 (0.0\%) & 0/42 (0.0\%)\\
Non-no-op actions/episode & 4.48 & 8.94\\
Mean redispatch $\ell_1$ (MW) & 31.53 & 77.25\\
Illegal/ambiguous selections & 0/0 & 0/0\\
\bottomrule
\end{tabularx}
\end{table}

\FloatBarrier

\begin{figure}[pos=ht!]
\vspace{-0.4cm}
\centering
\includegraphics[width=0.85\linewidth]{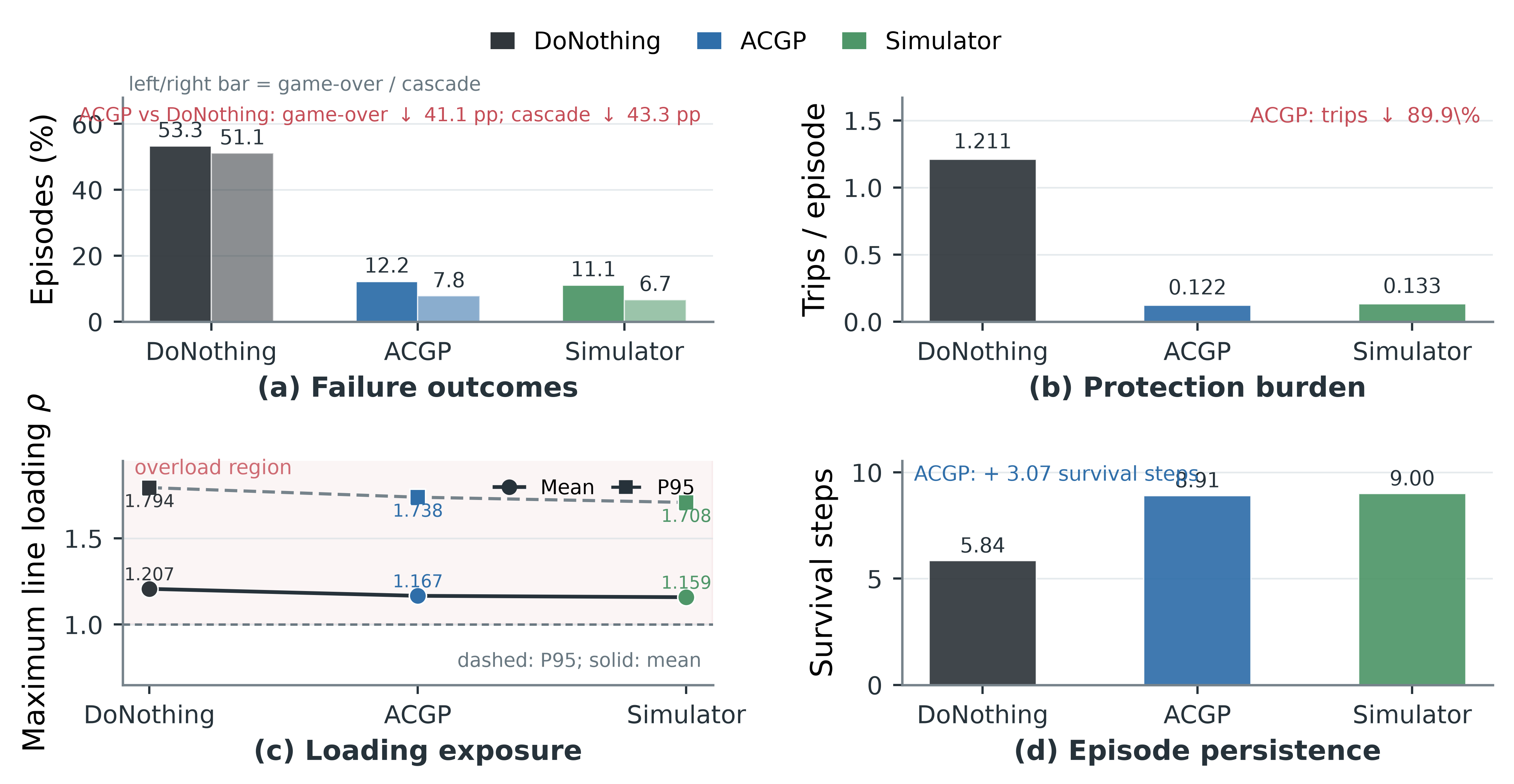}
\vspace{-0.2cm}
\caption{Aggregate closed-loop outcomes on 90 held-out episodes. The four panels report failure outcomes, automatic protection burden, loading exposure, and episode persistence for do-nothing control, ACGP, and the simulator reference.}
\label{fig:aggregateoutcomes}
\vspace{-0.8cm}
\end{figure}

Figure~\ref{fig:aggregateoutcomes} summarizes the episode-level closed-loop performance in terms of safety outcomes, protection activity, loading, and survival. ACGP closely approaches the simulator reference in terminal-failure and cascade rates while requiring fewer interventions. Table~\ref{tab:rescue-selectivity} further quantifies rescue capability, cascade prevention, and action selectivity. ACGP rescues 37 of 48 blackout-prone episodes and prevents cascades in 39 of 46 cascade-prone episodes, while none of the 42 episodes that survive under do-nothing operation are driven to terminal failure. The learned controller also requires approximately half as many non-no-op actions as the simulator reference and substantially less redispatch effort. The simulator-reference planner evaluates the same finite candidate set using true Grid2Op rollouts and therefore serves as an evaluation benchmark rather than a globally optimal controller. The relatively small performance gap further highlights the benefit of receding-horizon feedback, since only the selected action is executed and each subsequent decision is made from the newly observed grid state.
\subsection{Representative closed-loop cases}

\noindent\textbf{Blackout rescue.}

\begin{figure}[pos=ht!]
\centering
\includegraphics[width=0.6\linewidth]{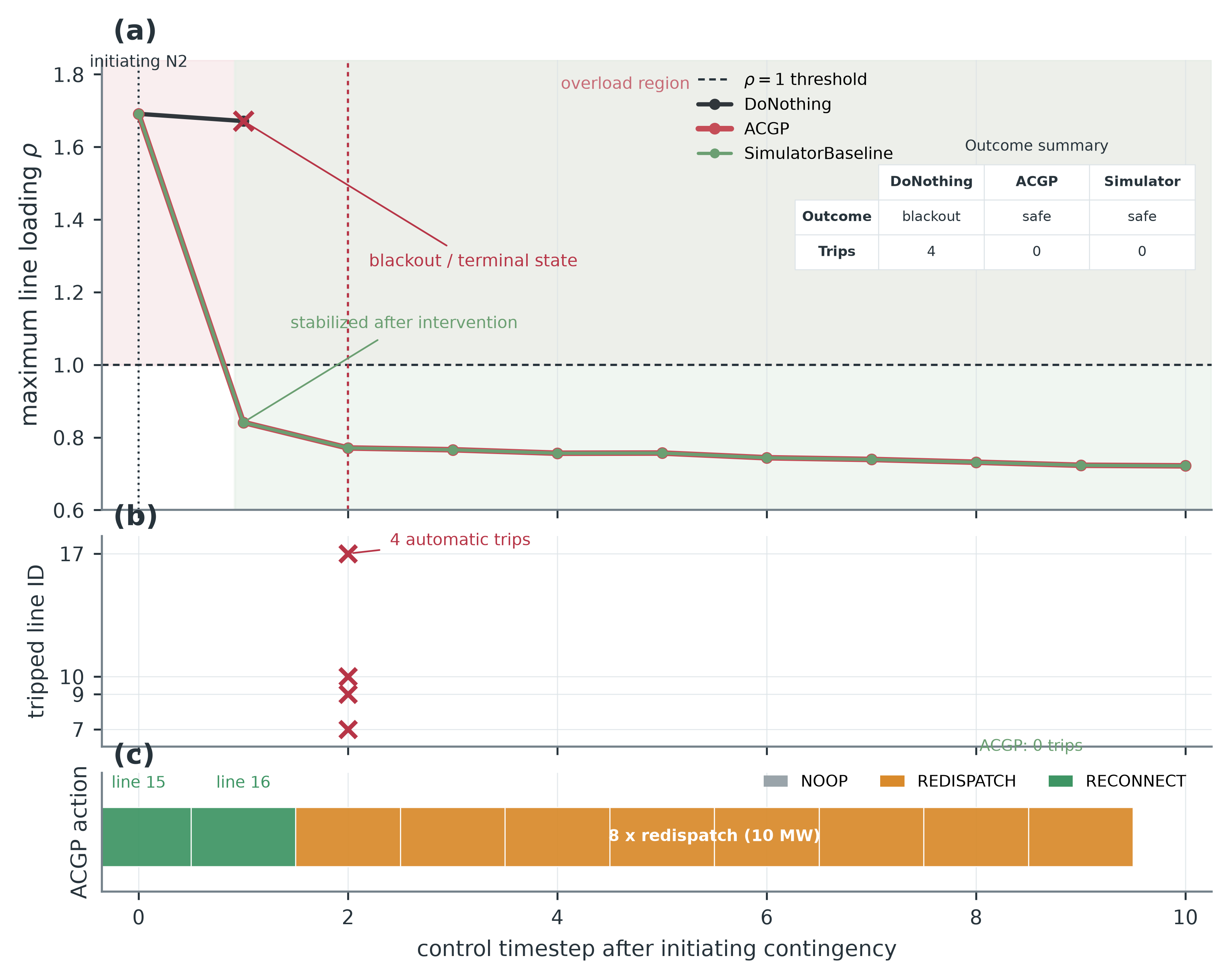}
\caption{Blackout-rescue trajectory under closed-loop control. Panel (a) compares maximum line loading for do-nothing operation, ACGP, and the simulator reference; panel (b) reports the automatic protection trips by line ID; panel (c) shows the aligned reconnection and redispatch actions.}
\label{fig:blackoutrescue}
\end{figure}

Figure~\ref{fig:blackoutrescue} presents a representative blackout-rescue episode. Under do-nothing operation, the system reaches terminal failure after four automatic protection trips. ACGP first reconnects lines 15 and 16 and then applies redispatch, after which the controlled trajectory remains below the overload threshold. This case demonstrates that interrupting cascade propagation may require coordinated corrective actions over successive control steps rather than a single isolated intervention.
\FloatBarrier

\FloatBarrier

\begin{figure}[pos=ht!]
\centering
\includegraphics[width=0.88\linewidth]{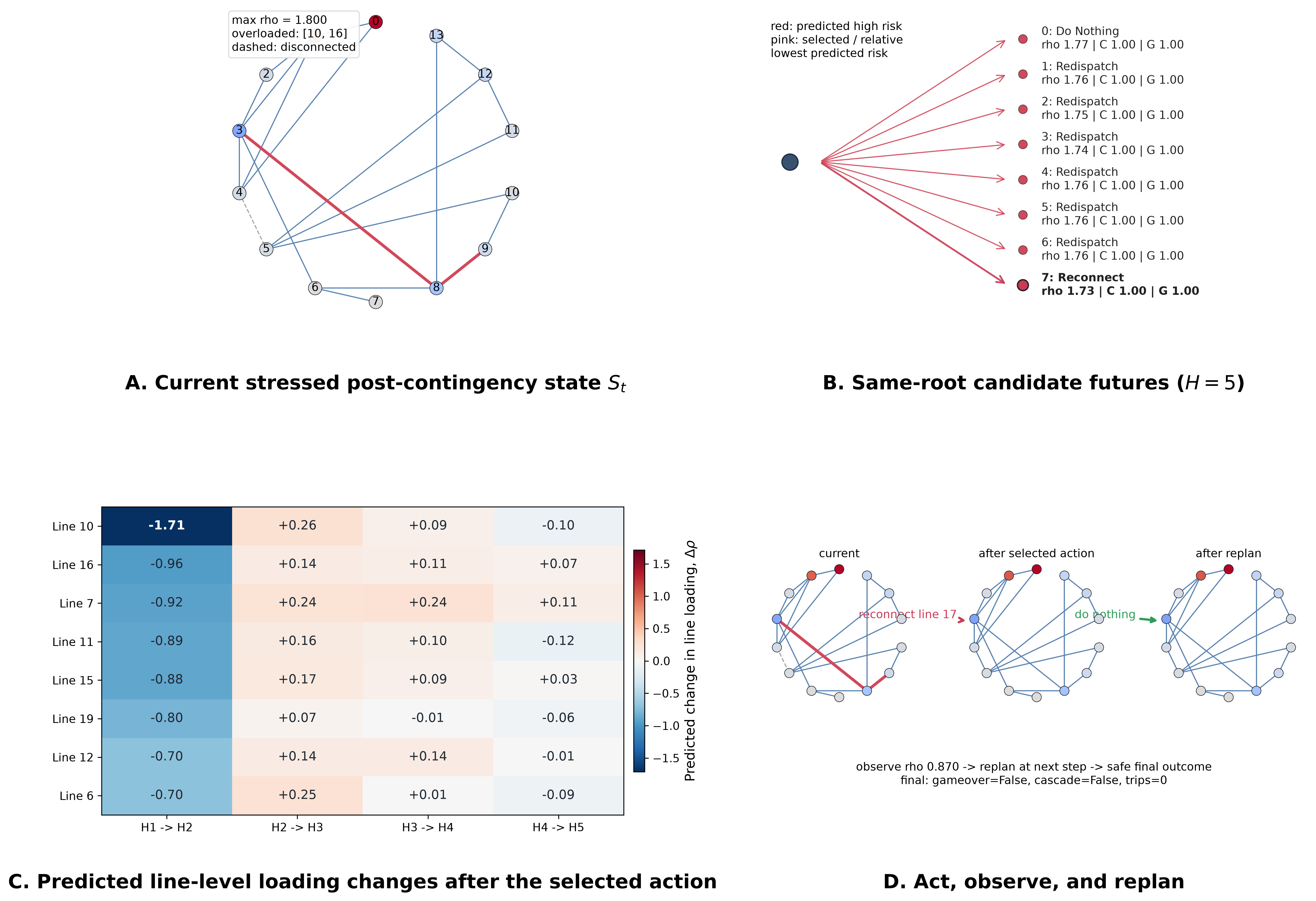}
\caption{Representative closed-loop decision using ACGP. Legal candidate futures are compared, a line-reconnection action is selected, the physical response is observed, and the controller replans at the next timestep.}
\label{fig:caseFmain}
\end{figure}

% \noindent\textbf{Representative feedback decision.}
Figure~\ref{fig:caseFmain} presents an N-1 episode with a current maximum line loading of $\rho=1.800$ and eight legal candidate actions. ACGP selects reconnection of line 17, which the simulator confirms as the safest retained candidate. Do-nothing operation and all six redispatch candidates lead to cascading failure, whereas the selected reconnection remains safe with a five-step maximum loading of 0.870. After executing this action, the controller observes the resulting physical state and replans, selecting do nothing at the subsequent decision step.

\vspace{-0.3cm}
\FloatBarrier

\begin{figure}[pos=ht!]
\centering
\vspace{-0.4cm}
\includegraphics[width=0.71\linewidth]{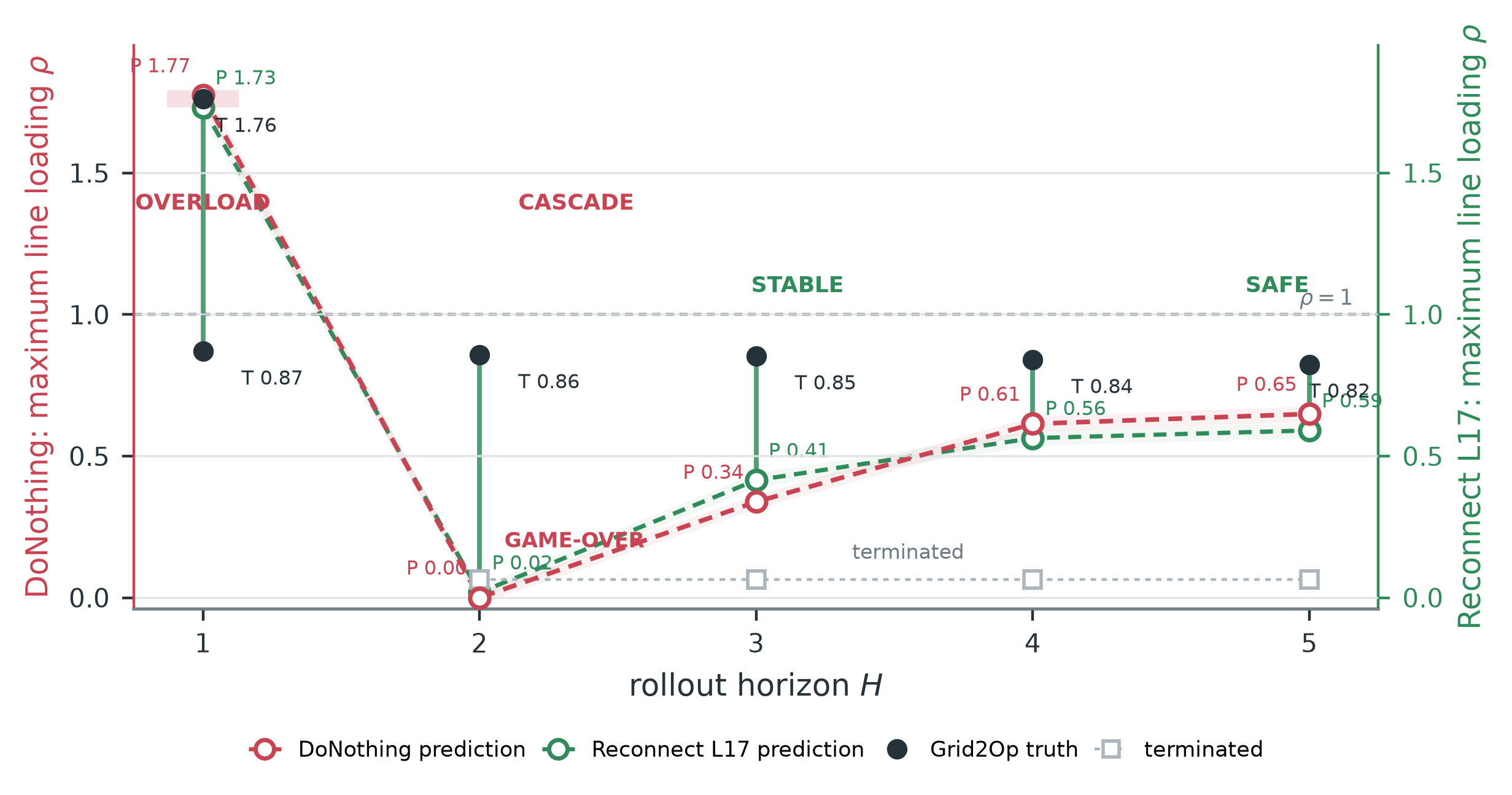}
\vspace{-0.3cm}
\caption{Five-step ACGP predictions and Grid2Op outcomes for do nothing and the selected intervention in the representative feedback case.}
\label{fig:caseFpred}
\end{figure}

Figure~\ref{fig:caseFpred} compares the five-step predicted and simulated trajectories for do-nothing operation and the selected reconnection. In this highly stressed condition, ACGP overestimates the absolute risk but still preserves the action ordering relevant to control. The selected reconnection is clearly separated from the failing do-nothing trajectory, showing that reliable relative ranking can be maintained even when the predicted risk magnitude is imperfect.

\FloatBarrier

\begin{figure}[pos=ht!]
\centering
\includegraphics[width=0.95\linewidth]{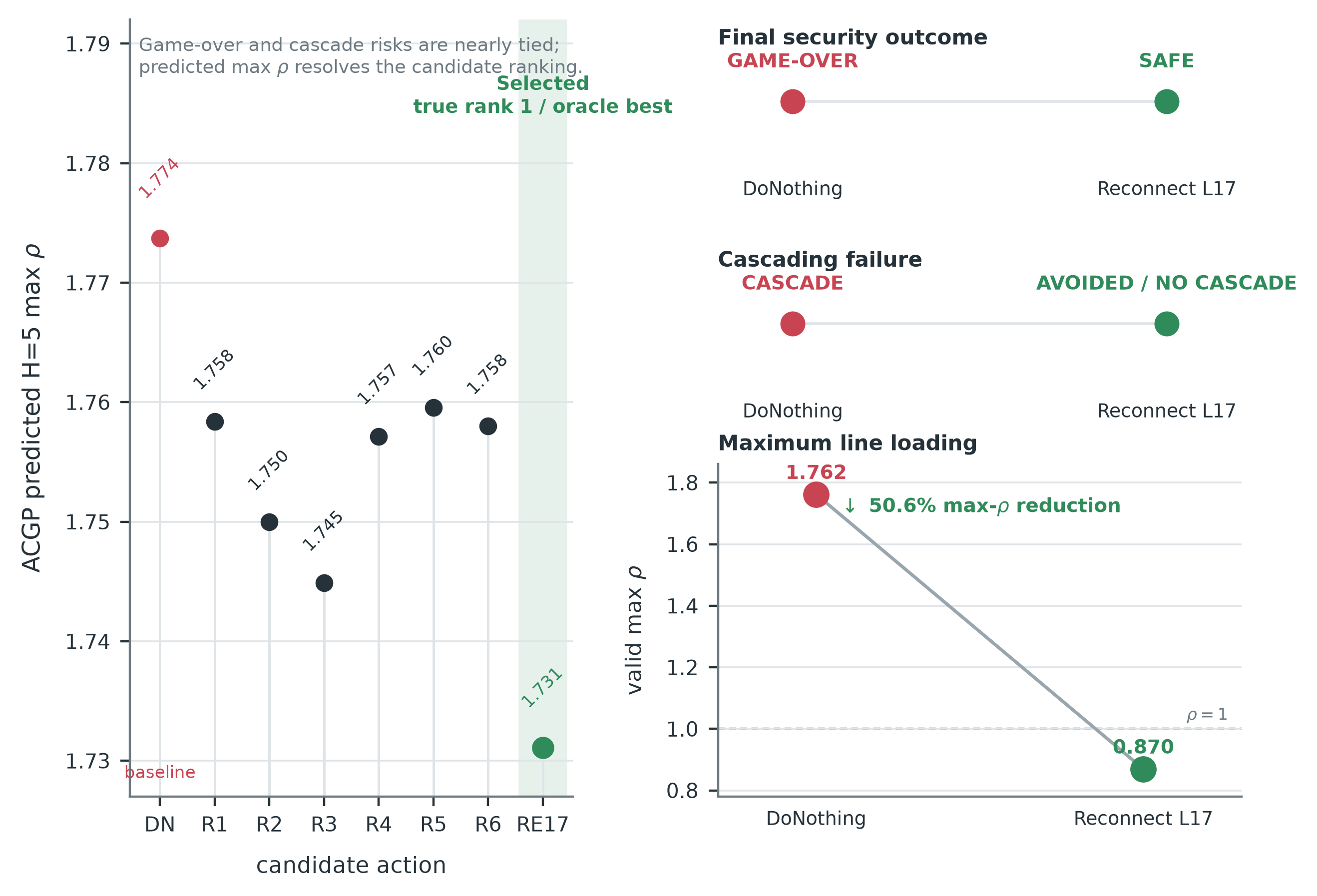}
\caption{Candidate-action landscape at the representative Case F decision step. Rank~1 denotes the safest action according to the simulator rollout.}
\label{fig:caseFlandscape}
\end{figure}

Figure~\ref{fig:caseFlandscape} and Table~\ref{tab:casef-candidates} provide the corresponding candidate-level comparison. The results show that the do-nothing and redispatch candidates are tightly clustered, whereas reconnection of line 17 is clearly identified as the safest retained candidate. The simulator confirms this action as true rank~1. This case highlights the intended use of ACGP in decision support. Even when the absolute multi-step predictions remain conservative, accurate relative ranking can still guide the controller toward an effective remedial action. Consequently, residual pointwise prediction error does not necessarily lead to control failure, provided that the safety-relevant ordering of candidate actions is preserved.
\vspace{-0.5cm}
\FloatBarrier

% \begin{table}[pos=ht!]
% \caption{Candidate-action comparison at the representative Case F decision step.}
% \label{tab:casef-candidates}
% \centering
% \small
% \begin{tabular}{@{}clccc@{}}
% \toprule
% ID & Action & Pred. max. $\rho$ & True max. $\rho$ & True rank\\
% \midrule
% 0 & Do nothing & 1.774 & 1.762 & 5\\
% 1 & Redispatch 1 & 1.758 & 1.762 & 6\\
% 2 & Redispatch 2 & 1.750 & 1.762 & 4\\
% 3 & Redispatch 3 & 1.745 & 1.762 & 3\\
% 4 & Redispatch 4 & 1.757 & 1.762 & 7\\
% 5 & Redispatch 5 & 1.760 & 1.762 & 2\\
% 6 & Redispatch 6 & 1.758 & 1.762 & 8\\
% 7 & Reconnect line 17 & \textbf{1.731} & \textbf{0.870} & \textbf{1}\\
% \bottomrule
% \end{tabular}

% % \footnotesize True ranks use the full simulator safety tuple and unrounded values; displayed loadings are rounded.
% \end{table}
% \FloatBarrier

\begin{table}[pos=ht!]
\caption{Candidate-action comparison at the representative decision step.}
\label{tab:casef-candidates}
\centering
\small
\begin{tabularx}{\linewidth}{
@{}>{\centering\arraybackslash}p{0.07\linewidth}
>{\raggedright\arraybackslash}X
>{\centering\arraybackslash}p{0.19\linewidth}
>{\centering\arraybackslash}p{0.18\linewidth}
>{\centering\arraybackslash}p{0.13\linewidth}@{}
}
\toprule
ID & Action & Pred. max. $\rho$ & True max. $\rho$ & True rank\\
\midrule
0 & Do nothing & 1.774 & 1.762 & 5\\
1 & Redispatch 1 & 1.758 & 1.762 & 6\\
2 & Redispatch 2 & 1.750 & 1.762 & 4\\
3 & Redispatch 3 & 1.745 & 1.762 & 3\\
4 & Redispatch 4 & 1.757 & 1.762 & 7\\
5 & Redispatch 5 & 1.760 & 1.762 & 2\\
6 & Redispatch 6 & 1.758 & 1.762 & 8\\
7 & Reconnect line 17 & \textbf{1.731} & \textbf{0.870} & \textbf{1}\\
\bottomrule
\end{tabularx}
\end{table}

\noindent\textbf{Cascade interruption and topology evolution.}

\begin{figure}[pos=ht!]
\centering
\includegraphics[width=0.8\linewidth]{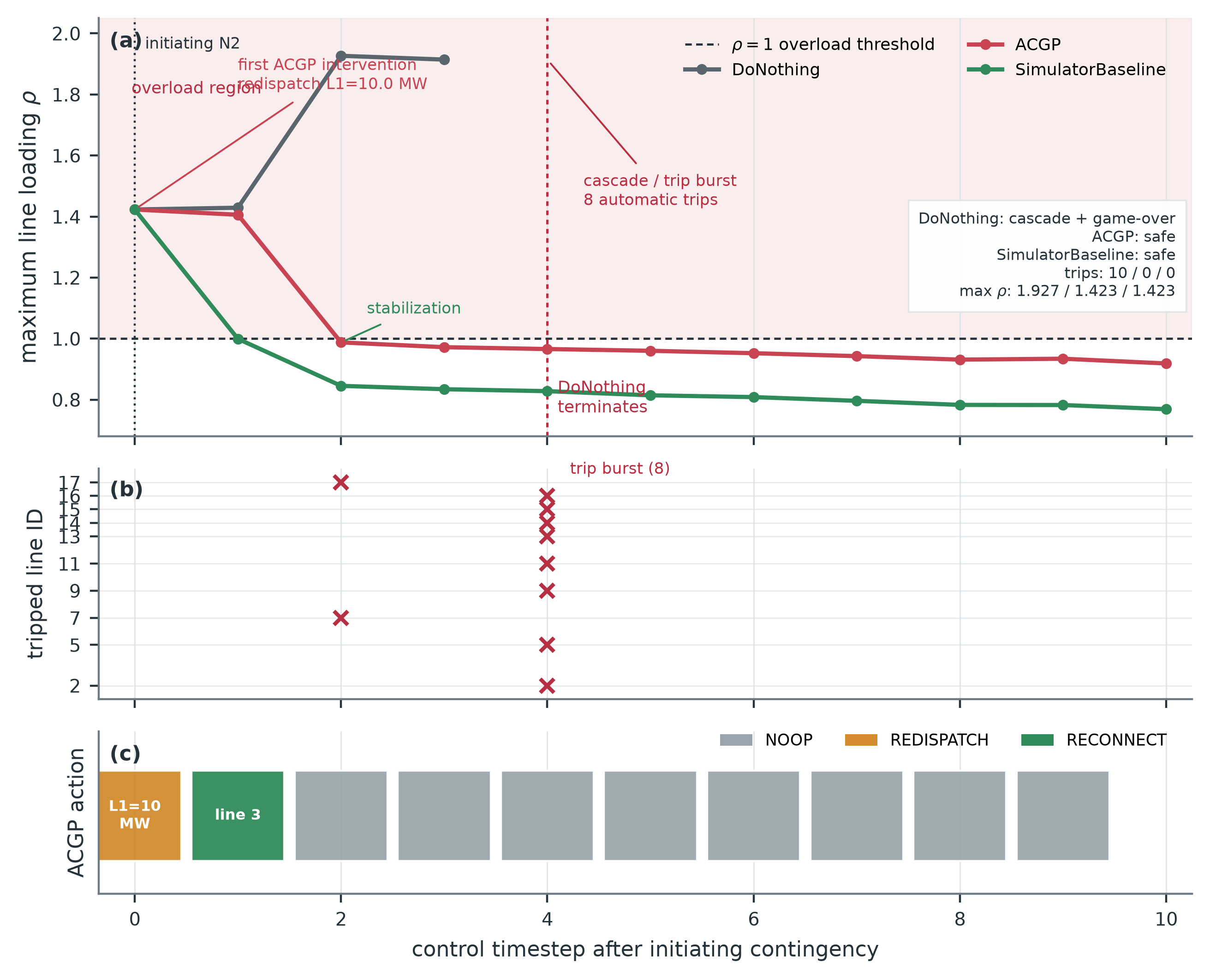}
\caption{Cascade interruption under closed-loop control. Do-nothing operation produces ten automatic trips, whereas ACGP suppresses the cascade.}
\label{fig:caseBtimeline}
\end{figure}

\begin{figure}[pos=ht!]
\centering
\begin{minipage}[t]{0.32\linewidth}
\centering
\includegraphics[width=\linewidth]{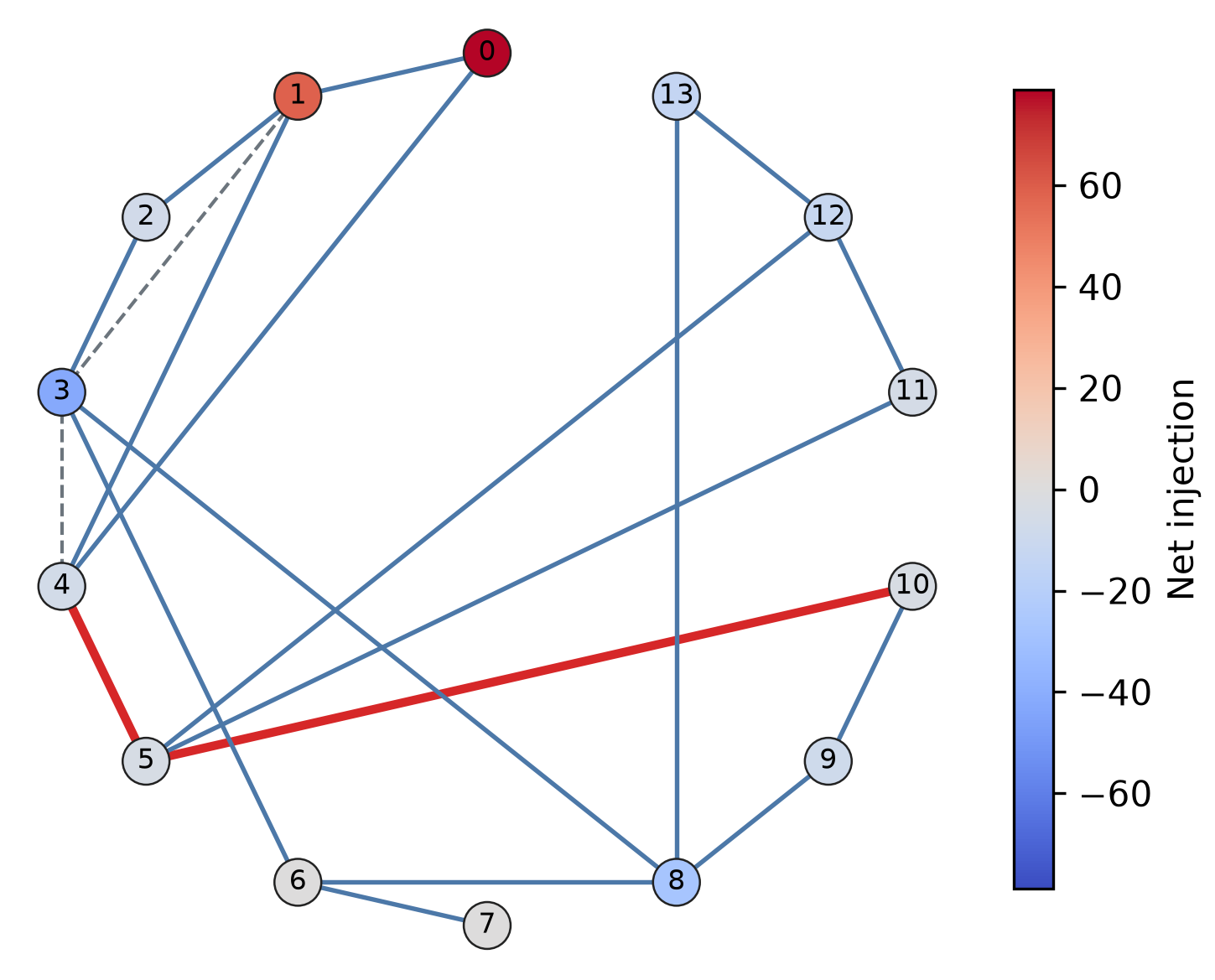}\\[-0.3em]
\textbf{(a)} Before intervention
\end{minipage}\hfill
\begin{minipage}[t]{0.32\linewidth}
\centering
\includegraphics[width=\linewidth]{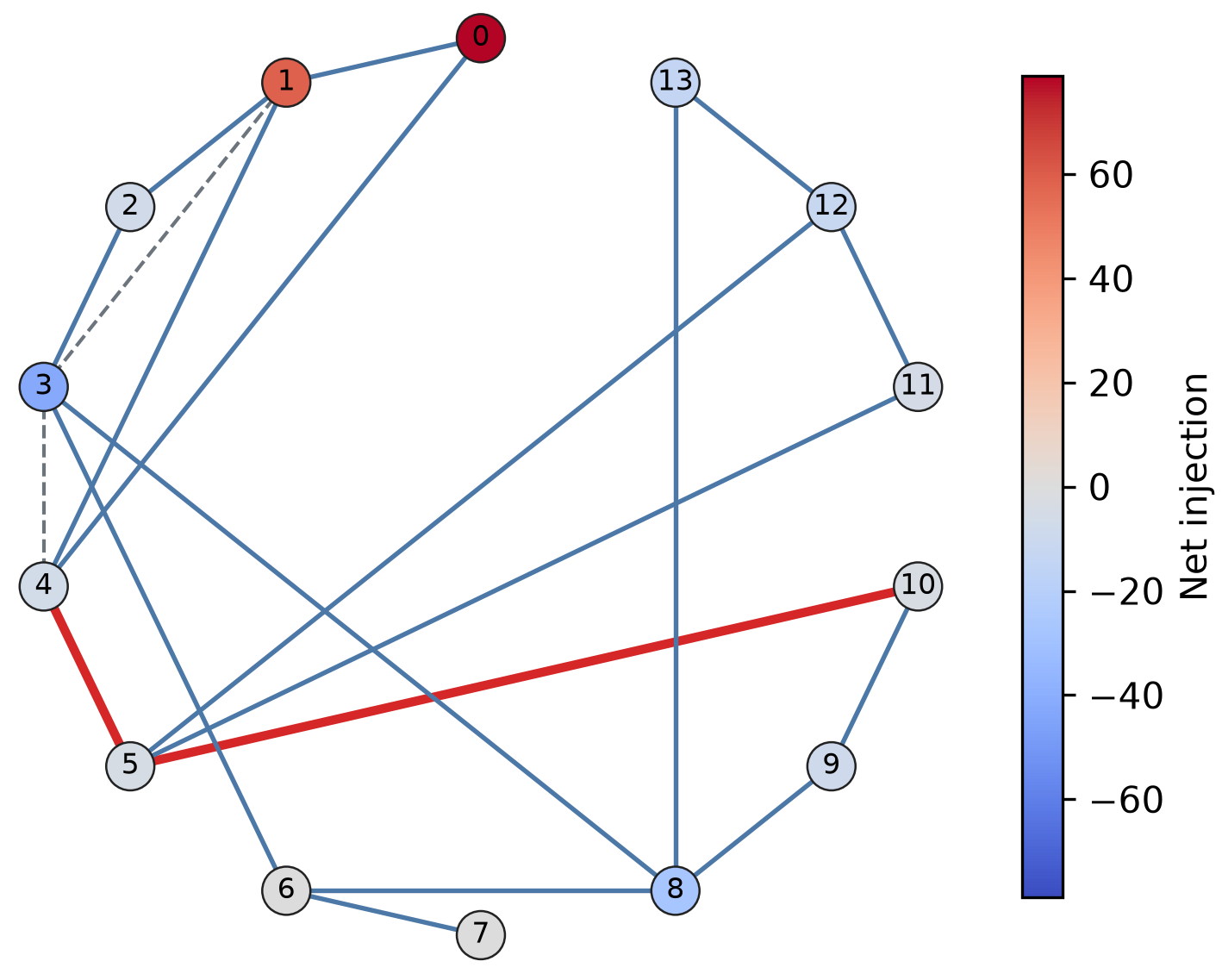}\\[-0.3em]
\textbf{(b)} Key action
\end{minipage}\hfill
\begin{minipage}[t]{0.32\linewidth}
\centering
\includegraphics[width=\linewidth]{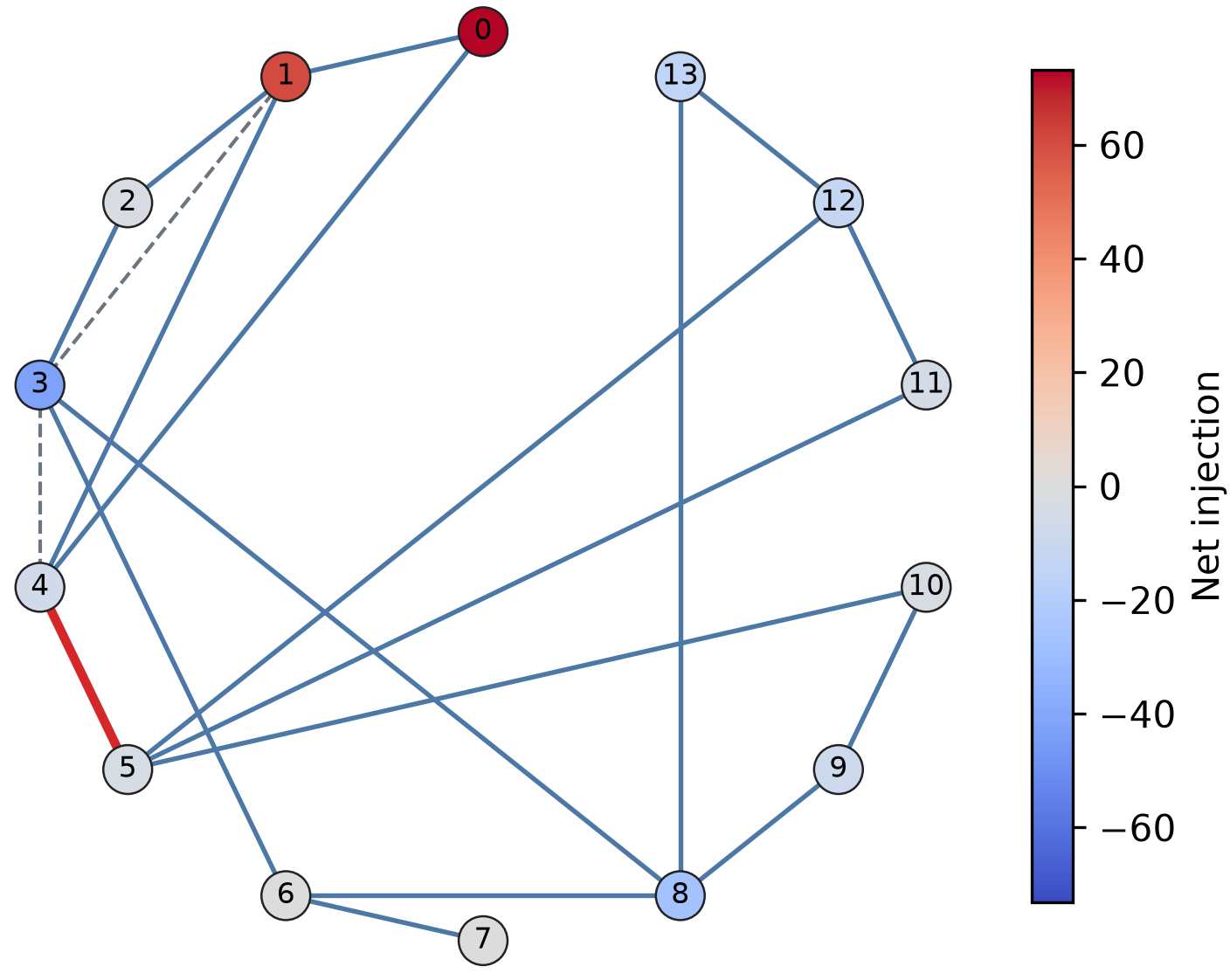}\\[-0.3em]
\textbf{(c)} Stabilized state
\end{minipage}
\caption{Topology sequence around the cascade-interruption intervention: (a) stressed pre-action state, (b) key planner intervention, and (c) stabilized post-action state. Dashed edges denote disconnected lines and node color indicates net injection.}
\label{fig:caseBgrid}
\end{figure}

Figure~\ref{fig:caseBtimeline} presents a second representative episode. Under do-nothing operation, the system experiences ten automatic line trips and eventually reaches terminal failure, whereas ACGP suppresses the propagation of the trip sequence. Following the early corrective actions, the controlled trajectory returns below the overload threshold. Figure~\ref{fig:caseBgrid} shows the corresponding network states before, during, and after the key intervention. Because the node and edge layout is kept fixed across the three panels, the physical effect of the corrective action can be seen directly. Network connectivity is modified before the protection sequence propagates further, and the post-action state remains stable without additional automatic trips.

\FloatBarrier

\FloatBarrier

\begin{figure}[pos=ht!]
\centering
\begin{minipage}[t]{0.32\linewidth}
\centering
\includegraphics[width=\linewidth]{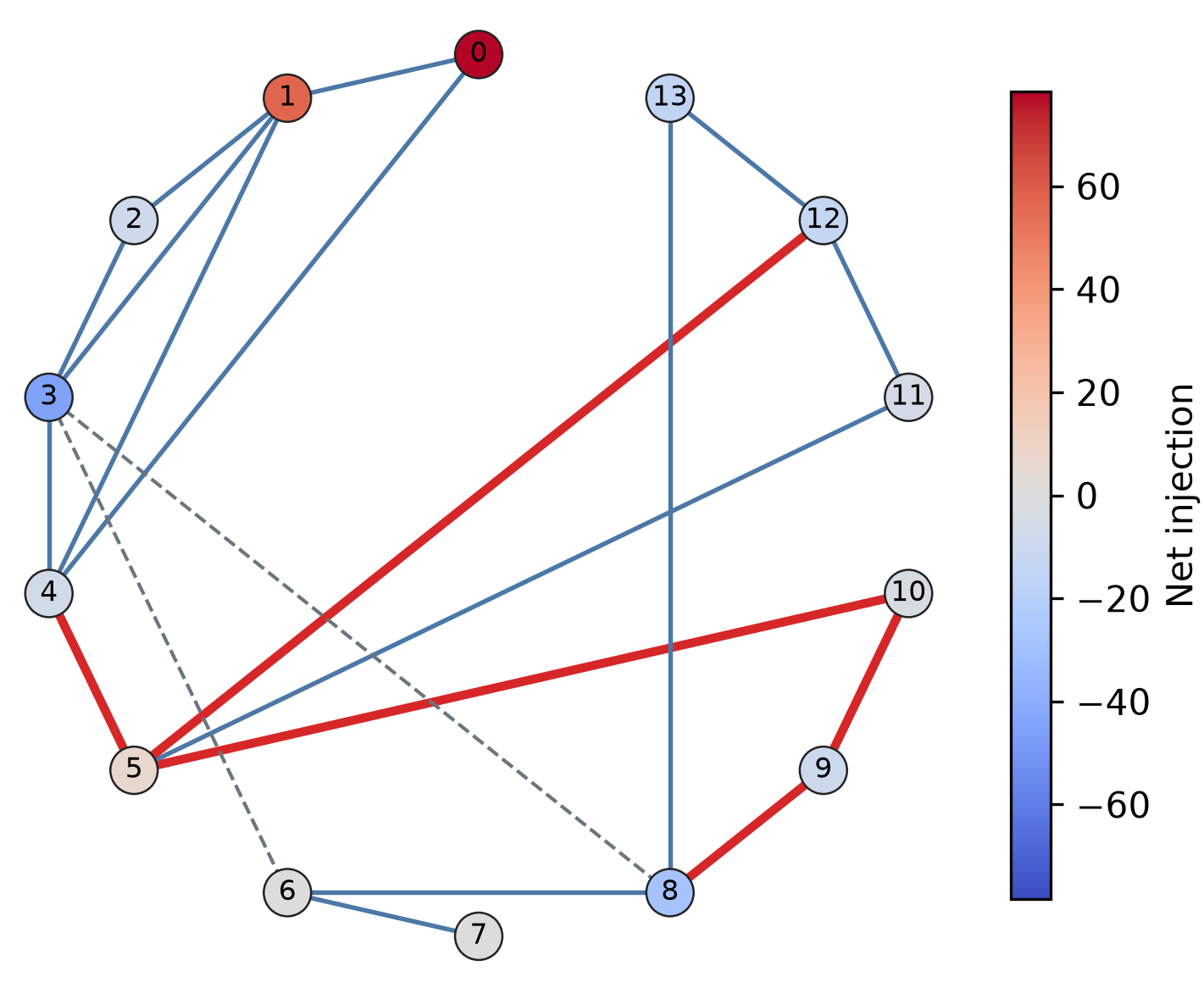}\\[-0.3em]
\textbf{(a)} Before intervention
\end{minipage}\hfill
\begin{minipage}[t]{0.32\linewidth}
\centering
\includegraphics[width=\linewidth]{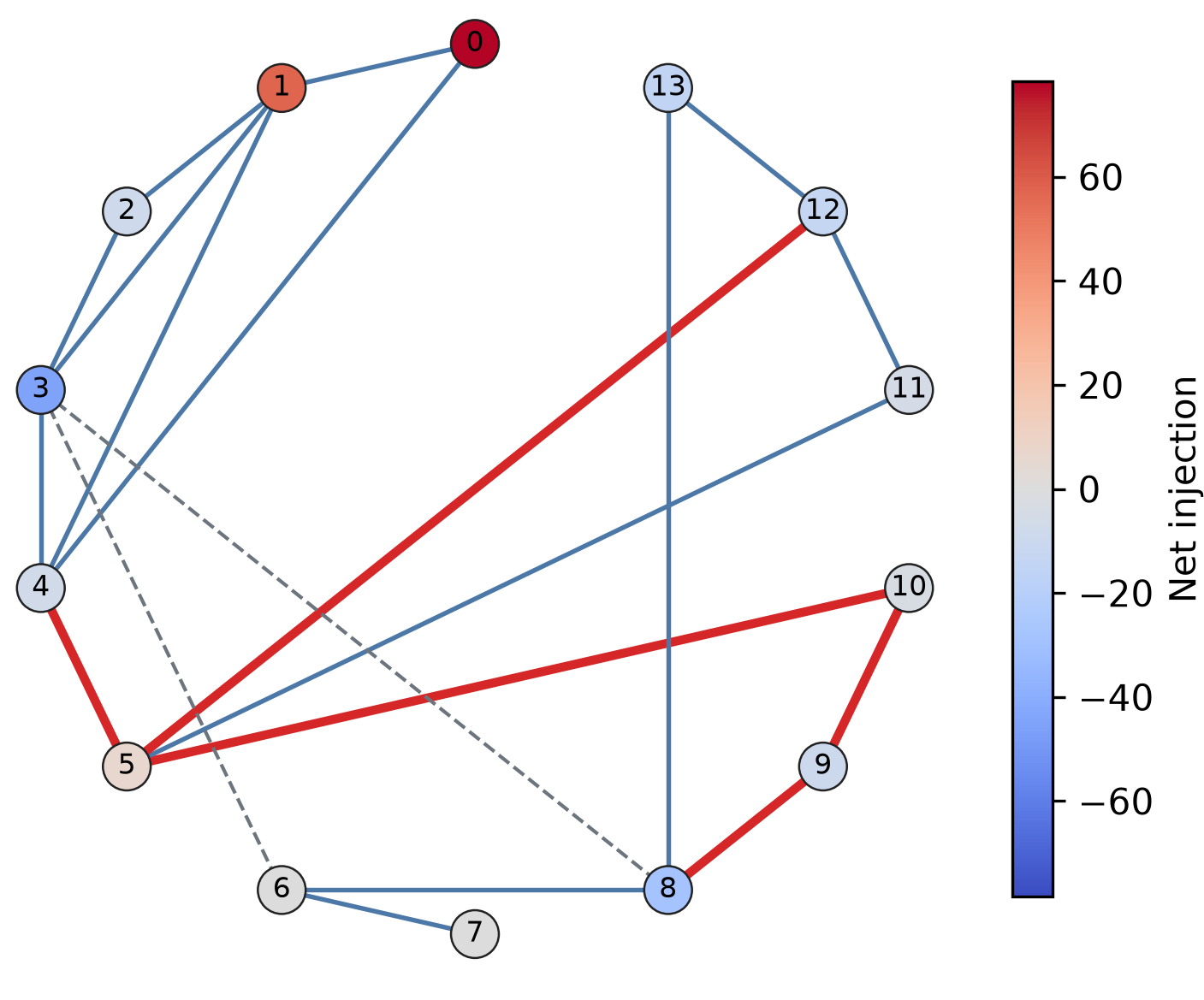}\\[-0.3em]
\textbf{(b)} Key action
\end{minipage}\hfill
\begin{minipage}[t]{0.32\linewidth}
\centering
\includegraphics[width=\linewidth]{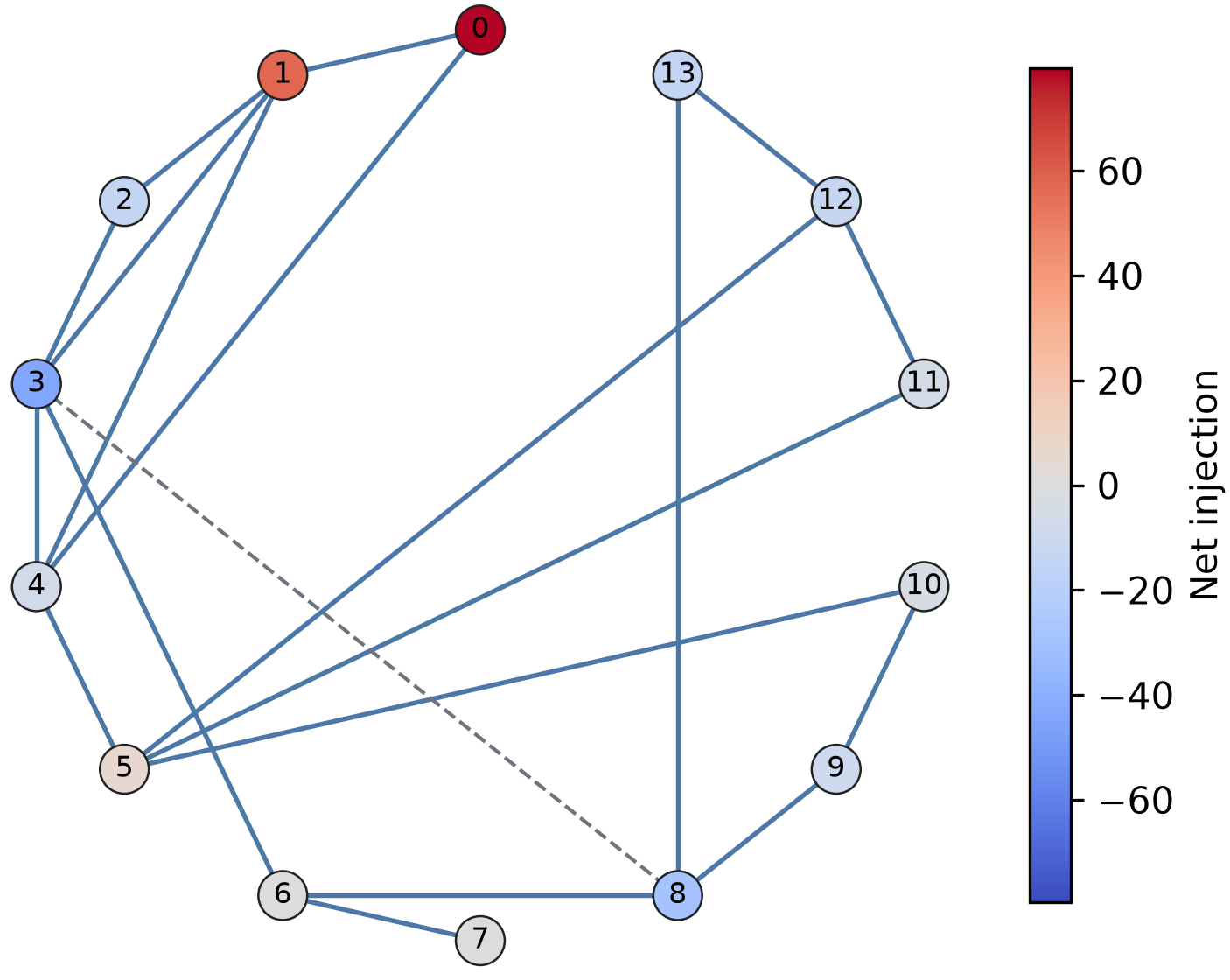}\\[-0.3em]
\textbf{(c)} Stabilized state
\end{minipage}
\caption{Topology sequence for the blackout-rescue intervention: (a) stressed pre-action state, (b) corrective action, and (c) stabilized post-action state. Dashed edges denote disconnected lines and node color indicates net injection.}
\label{fig:caseAgrid}
\end{figure}

Figure~\ref{fig:caseAgrid} presents an additional held-out topology sequence from the blackout-rescue setting. It provides complementary physical evidence that the selected interventions produce meaningful changes in network configuration, consistent with the improved closed-loop safety outcomes.

\FloatBarrier

% Approximately 54.6\% of ACGP-guided decisions differ from the simulator-ranked best action, while the overall terminal-failure rate remains close to that of the simulator reference. The subsequent terminal-failure frequency is 2.05\% following a non-optimal selection and 0.82\% following a simulator-optimal selection. Combined with the Top-3 recovery rate of 0.8801, these results indicate that the candidate set frequently contains multiple actions with comparable safety performance. Consequently, closed-loop mitigation does not require exact recovery of the simulator-ranked Top-1 action at every decision step, as long as the controller preserves the safety-relevant ordering among candidate actions and updates its decision using the subsequently observed system state.
The aggregate and case-study results consistently indicate that action-conditioned multi-step prediction can provide effective support for closed-loop cascading-failure mitigation. The selected remedial actions alter the subsequent loading evolution, network connectivity, and protection response, while receding-horizon replanning continuously incorporates the observed system state into the control process. This combination enables safety-oriented comparison of legal corrective actions without requiring exact reproduction of the complete future cascade trajectory.

% ==================================================================
\FloatBarrier
\section{Conclusion}
\label{sec:conclusion}
% ==================================================================

This paper developed an Action-Conditioned Graph Predictor (ACGP) for multi-step cascading-failure prediction and closed-loop remedial control. Same-state counterfactual branching enables alternative legal redispatch and reconnection actions to be evaluated under identical operating and exogenous conditions, thereby isolating their effects on subsequent system evolution. ACGP captures the resulting changes in electrical states, line connectivity, protection actions, and terminal risk, and is integrated into a receding-horizon controller that repeatedly evaluates candidate trajectories and updates its decision from the subsequently observed grid state. On the held-out counterfactual test set, ACGP achieves a five-step line-loading MAE of 0.0369 and a redispatch pairwise ranking accuracy of 0.806. Action-input ablations reduce the ranking accuracy to approximately 0.5, demonstrating that the learned same-root ordering depends on the applied remedial action rather than on the stressed operating state alone. In closed-loop evaluation, ACGP reduces the terminal-failure rate from 53.3\% to 12.2\% and the cascade rate from 51.1\% to 7.8\%, approaching the simulator-reference values of 11.1\% and 6.7\%, respectively. The case studies further indicate that successful cascade mitigation does not require exact recovery of the simulator-ranked best action at every decision step. Preserving the safety-relevant ordering among candidate actions, together with repeated state feedback and replanning, can provide effective control even when absolute multi-step predictions remain imperfect.

The present study is limited to the Grid2Op \texttt{l2rpn\_case14\_sandbox} environment, a five-step prediction horizon, and a finite remedial-action set consisting of do nothing, redispatch, and line reconnection. The simulation-based evaluation does not yet establish scalability or robustness under larger transmission networks, broader topology-control actions, measurement and model uncertainty, communication delays, varying protection settings, or out-of-distribution operating conditions. Future work will extend the framework to larger benchmark systems and richer corrective-action spaces, while incorporating uncertainty-aware prediction and ranking and evaluating robustness under operating conditions that differ from the training distribution.
% ==================================================================
\section*{CRediT authorship contribution statement}
Fangxin Liu: Conceptualization, methodology, investigation, data curation, formal analysis, visualization, and writing -- original draft. Zhentao Tian: Data curation, visualization, validation, and writing -- review and editing. Both authors reviewed and approved the final manuscript.

\section*{Declaration of competing interest}
The authors declare that they have no known competing financial interests or personal relationships that could have appeared to influence the work reported in this paper.

\section*{Funding}
This research did not receive any specific grant from funding agencies in the public, commercial, or not-for-profit sectors.

\section*{Data availability}
The public \texttt{l2rpn\_case14\_sandbox} benchmark environment is available through Grid2Op. The counterfactual datasets, trained checkpoints, and evaluation outputs reported here are generated by the reproducible experimental pipeline described in this manuscript.

% ==================================================================

% ==================================================================
%% Numbered Elsevier CAS bibliography style.
\bibliographystyle{model1-num-names}
\bibliography{references}

\begin{thebibliography}{29}
\expandafter\ifx\csname natexlab\endcsname\relax\def\natexlab#1{#1}\fi
\providecommand{\url}[1]{\texttt{#1}}
\providecommand{\href}[2]{#2}
\providecommand{\path}[1]{#1}
\providecommand{\DOIprefix}{doi:}
\providecommand{\ArXivprefix}{arXiv:}
\providecommand{\URLprefix}{URL: }
\providecommand{\Pubmedprefix}{pmid:}
\providecommand{\doi}[1]{\href{http://dx.doi.org/#1}{\path{#1}}}
\providecommand{\Pubmed}[1]{\href{pmid:#1}{\path{#1}}}
\providecommand{\bibinfo}[2]{#2}
\ifx\xfnm\relax \def\xfnm[#1]{\unskip,\space#1}\fi
%Type = Article
\bibitem[{Li et~al.(2025)Li, Liu, Yin, Jiang, Zhang, and Liu}]{li2025study}
\bibinfo{author}{B.~Li}, \bibinfo{author}{C.~Liu}, \bibinfo{author}{Y.~Yin}, \bibinfo{author}{Q.~Jiang}, \bibinfo{author}{Y.~Zhang}, \bibinfo{author}{T.~Liu},
\newblock \bibinfo{title}{Study on power system resilience assessment considering cascading failures during wildfire disasters},
\newblock \bibinfo{journal}{Energy Reports} \bibinfo{volume}{13} (\bibinfo{year}{2025}) \bibinfo{pages}{1819--1833}.
%Type = Article
\bibitem[{Zakariya and Teh(2023)}]{zakariya2023systematic}
\bibinfo{author}{M.~Zakariya}, \bibinfo{author}{J.~Teh},
\newblock \bibinfo{title}{A systematic review on cascading failures models in renewable power systems with dynamics perspective and protections modeling},
\newblock \bibinfo{journal}{Electric Power Systems Research} \bibinfo{volume}{214} (\bibinfo{year}{2023}) \bibinfo{pages}{108928}.
%Type = Article
\bibitem[{Pei et~al.(2026)Pei, Yang, and Xu}]{pei2026risk}
\bibinfo{author}{Y.~Pei}, \bibinfo{author}{J.~Yang}, \bibinfo{author}{J.~Xu},
\newblock \bibinfo{title}{Risk analysis of cascading failures considering the relay protection characteristics in wind/photovoltaic integrated power systems},
\newblock \bibinfo{journal}{Electrical Engineering} \bibinfo{volume}{108} (\bibinfo{year}{2026}) \bibinfo{pages}{279}.
%Type = Article
\bibitem[{Liu et~al.(2021)Liu, Zhang, and Tse}]{9272673}
\bibinfo{author}{D.~Liu}, \bibinfo{author}{X.~Zhang}, \bibinfo{author}{C.~K. Tse},
\newblock \bibinfo{title}{A tutorial on modeling and analysis of cascading failure in future power grids},
\newblock \bibinfo{journal}{IEEE Transactions on Circuits and Systems II: Express Briefs} \bibinfo{volume}{68} (\bibinfo{year}{2021}) \bibinfo{pages}{49--55}.
%Type = Article
\bibitem[{Li et~al.(2023)Li, Tse, Liu, and Zhang}]{li2023cascading}
\bibinfo{author}{M.~J. Li}, \bibinfo{author}{C.~K. Tse}, \bibinfo{author}{D.~Liu}, \bibinfo{author}{X.~Zhang},
\newblock \bibinfo{title}{Cascading failure propagation and mitigation strategies in power systems},
\newblock \bibinfo{journal}{IEEE Systems Journal} \bibinfo{volume}{17} (\bibinfo{year}{2023}) \bibinfo{pages}{3282--3293}.
%Type = Article
\bibitem[{Zhao et~al.(2025)Zhao, Cai, Cozzani, and Liu}]{zhao2025failure}
\bibinfo{author}{Y.~Zhao}, \bibinfo{author}{B.~Cai}, \bibinfo{author}{V.~Cozzani}, \bibinfo{author}{Y.~Liu},
\newblock \bibinfo{title}{Failure dependence and cascading failures: A literature review and research opportunities},
\newblock \bibinfo{journal}{Reliability Engineering \& System Safety} \bibinfo{volume}{256} (\bibinfo{year}{2025}) \bibinfo{pages}{110766}.
%Type = Article
\bibitem[{Hayat et~al.(2025)Hayat, Adnan, Iqbal, Mohamed, and Tariq}]{hayat2025expert}
\bibinfo{author}{F.~Hayat}, \bibinfo{author}{M.~Adnan}, \bibinfo{author}{M.~S. Iqbal}, \bibinfo{author}{S.~E.~G. Mohamed}, \bibinfo{author}{M.~Tariq},
\newblock \bibinfo{title}{Expert and intelligent systems for assessment and mitigation of cascading failures in smart grids: Research challenges and survey},
\newblock \bibinfo{journal}{Results in Engineering}  (\bibinfo{year}{2025}) \bibinfo{pages}{107148}.
%Type = Article
\bibitem[{Li and Tse(2024)}]{10497182}
\bibinfo{author}{M.~J. Li}, \bibinfo{author}{C.~K. Tse},
\newblock \bibinfo{title}{Quantification of cascading failure propagation in power systems},
\newblock \bibinfo{journal}{IEEE Transactions on Circuits and Systems I: Regular Papers} \bibinfo{volume}{71} (\bibinfo{year}{2024}) \bibinfo{pages}{3717--3725}.
%Type = Article
\bibitem[{He et~al.(2025)He, Zhou, Yang, Li, Liu, and Liu}]{10931178}
\bibinfo{author}{S.~He}, \bibinfo{author}{Y.~Zhou}, \bibinfo{author}{Y.~Yang}, \bibinfo{author}{X.~Li}, \bibinfo{author}{Y.~Liu}, \bibinfo{author}{T.~Liu},
\newblock \bibinfo{title}{Vulnerable sequence identification for sequential cascading failure analysis in power grid},
\newblock \bibinfo{journal}{IEEE Transactions on Industrial Informatics} \bibinfo{volume}{21} (\bibinfo{year}{2025}) \bibinfo{pages}{4830--4840}.
%Type = Article
\bibitem[{Deng et~al.(2025)Deng, Xie, Huang, Shen, and Pan}]{deng2025cascading}
\bibinfo{author}{H.~Deng}, \bibinfo{author}{P.~Xie}, \bibinfo{author}{H.~Huang}, \bibinfo{author}{J.~Shen}, \bibinfo{author}{D.~Pan},
\newblock \bibinfo{title}{Cascading fault prevention and control strategy based on economic dispatch of {AC/DC} systems},
\newblock \bibinfo{journal}{Electrical Engineering} \bibinfo{volume}{107} (\bibinfo{year}{2025}) \bibinfo{pages}{2551--2567}.
%Type = Article
\bibitem[{Li et~al.(2024)Li, Tse, and Yi}]{10294236}
\bibinfo{author}{M.~J. Li}, \bibinfo{author}{C.~K. Tse}, \bibinfo{author}{M.~Yi},
\newblock \bibinfo{title}{Steady-state cascading failure model with voltage instability event detection},
\newblock \bibinfo{journal}{IEEE Transactions on Circuits and Systems I: Regular Papers} \bibinfo{volume}{71} (\bibinfo{year}{2024}) \bibinfo{pages}{463--472}.
%Type = Article
\bibitem[{Wu et~al.(2021)Wu, Wu, and Modiano}]{9384281}
\bibinfo{author}{X.~Wu}, \bibinfo{author}{D.~Wu}, \bibinfo{author}{E.~Modiano},
\newblock \bibinfo{title}{Predicting failure cascades in large scale power systems via the influence model framework},
\newblock \bibinfo{journal}{IEEE Transactions on Power Systems} \bibinfo{volume}{36} (\bibinfo{year}{2021}) \bibinfo{pages}{4778--4790}.
%Type = Article
\bibitem[{Ma et~al.(2020)Ma, Jin, Eftekharnejad, Zafarani, and Philippe}]{9050711}
\bibinfo{author}{R.~Ma}, \bibinfo{author}{S.~Jin}, \bibinfo{author}{S.~Eftekharnejad}, \bibinfo{author}{R.~Zafarani}, \bibinfo{author}{W.~P.~J. Philippe},
\newblock \bibinfo{title}{A probabilistic cascading failure model for dynamic operating conditions},
\newblock \bibinfo{journal}{IEEE Access} \bibinfo{volume}{8} (\bibinfo{year}{2020}) \bibinfo{pages}{61741--61753}.
%Type = Article
\bibitem[{Al-Rousan et~al.(2022)Al-Rousan, Wang, and Lin}]{9733408}
\bibinfo{author}{W.~Al-Rousan}, \bibinfo{author}{C.~Wang}, \bibinfo{author}{F.~Lin},
\newblock \bibinfo{title}{A discrete-event system approach for modeling and mitigating power system cascading failures},
\newblock \bibinfo{journal}{IEEE Transactions on Control Systems Technology} \bibinfo{volume}{30} (\bibinfo{year}{2022}) \bibinfo{pages}{2547--2560}.
%Type = Article
\bibitem[{Liu et~al.(2025)Liu, Yan, Liu, and Chung}]{10849943}
\bibinfo{author}{T.~Liu}, \bibinfo{author}{J.~Yan}, \bibinfo{author}{Y.~Liu}, \bibinfo{author}{C.~Y. Chung},
\newblock \bibinfo{title}{Cascading failure screening based on gradient boosting decision tree for hvdc sending-end systems with high wind power penetration},
\newblock \bibinfo{journal}{IEEE Transactions on Power Systems} \bibinfo{volume}{40} (\bibinfo{year}{2025}) \bibinfo{pages}{3682--3694}.
%Type = Article
\bibitem[{Tang and Yue(2025)}]{tang2025deep}
\bibinfo{author}{X.~Tang}, \bibinfo{author}{P.~Yue},
\newblock \bibinfo{title}{Deep learning-based detection of cascading failure events in power systems using an event-triggered hybrid model framework},
\newblock \bibinfo{journal}{Electrical Engineering} \bibinfo{volume}{107} (\bibinfo{year}{2025}) \bibinfo{pages}{15025--15039}.
%Type = Article
\bibitem[{Liu et~al.(2021)Liu, Wu, Li, Shen, Qian, and Liu}]{9215038}
\bibinfo{author}{L.~Liu}, \bibinfo{author}{H.~Wu}, \bibinfo{author}{L.~Li}, \bibinfo{author}{D.~Shen}, \bibinfo{author}{F.~Qian}, \bibinfo{author}{J.~Liu},
\newblock \bibinfo{title}{Cascading failure pattern identification in power systems based on sequential pattern mining},
\newblock \bibinfo{journal}{IEEE Transactions on Power Systems} \bibinfo{volume}{36} (\bibinfo{year}{2021}) \bibinfo{pages}{1856--1866}.
%Type = Article
\bibitem[{Zhu et~al.(2023)Zhu, Zhou, Wei, and Wang}]{9881910}
\bibinfo{author}{Y.~Zhu}, \bibinfo{author}{Y.~Zhou}, \bibinfo{author}{W.~Wei}, \bibinfo{author}{N.~Wang},
\newblock \bibinfo{title}{Cascading failure analysis based on a physics-informed graph neural network},
\newblock \bibinfo{journal}{IEEE Transactions on Power Systems} \bibinfo{volume}{38} (\bibinfo{year}{2023}) \bibinfo{pages}{3632--3641}.
%Type = Article
\bibitem[{Varbella et~al.(2023)Varbella, Gjorgiev, and Sansavini}]{varbella2023geometric}
\bibinfo{author}{A.~Varbella}, \bibinfo{author}{B.~Gjorgiev}, \bibinfo{author}{G.~Sansavini},
\newblock \bibinfo{title}{Geometric deep learning for online prediction of cascading failures in power grids},
\newblock \bibinfo{journal}{Reliability Engineering \& System Safety} \bibinfo{volume}{237} (\bibinfo{year}{2023}) \bibinfo{pages}{109341}.
%Type = Article
\bibitem[{Sami and Naeini(2026)}]{11653266}
\bibinfo{author}{N.~M. Sami}, \bibinfo{author}{M.~Naeini},
\newblock \bibinfo{title}{Graph neural network framework for line status prediction in power systems cascading failures},
\newblock \bibinfo{journal}{IEEE Systems Journal}  (\bibinfo{year}{2026}) \bibinfo{pages}{1--4}.
%Type = Article
\bibitem[{Soleimani et~al.(2025)Soleimani, Afshar, and Atrianfar}]{soleimani2025critical}
\bibinfo{author}{S.~Soleimani}, \bibinfo{author}{A.~Afshar}, \bibinfo{author}{H.~Atrianfar},
\newblock \bibinfo{title}{Critical component analysis of cyber-physical power systems in cascading failures using graph convolutional networks: An energy-based approach},
\newblock \bibinfo{journal}{Sustainable Energy, Grids and Networks} \bibinfo{volume}{42} (\bibinfo{year}{2025}) \bibinfo{pages}{101653}.
%Type = Article
\bibitem[{Zhao et~al.(2026)Zhao, Li, Wu, Lei, Tan, and Li}]{zhao2026anomalous}
\bibinfo{author}{C.~Zhao}, \bibinfo{author}{B.~Li}, \bibinfo{author}{L.~Wu}, \bibinfo{author}{Z.~Lei}, \bibinfo{author}{Z.~Tan}, \bibinfo{author}{X.~Li},
\newblock \bibinfo{title}{Anomalous node localization and impact scope quantification in power system fault links using graph attention network with cascading failure propagation model},
\newblock \bibinfo{journal}{Journal of Engineering and Applied Science} \bibinfo{volume}{73} (\bibinfo{year}{2026}) \bibinfo{pages}{261}.
%Type = Article
\bibitem[{Zhang et~al.(2026)Zhang, Xie, Li et~al.}]{zhang2026egds}
\bibinfo{author}{J.~Zhang}, \bibinfo{author}{J.~Xie}, \bibinfo{author}{F.~Li}, et~al.,
\newblock \bibinfo{title}{Egds-cnn: a deep learning framework for critical node identification and vulnerability analysis of power systems},
\newblock \bibinfo{journal}{Electrical Engineering} \bibinfo{volume}{108} (\bibinfo{year}{2026}) \bibinfo{pages}{384}.
%Type = Article
\bibitem[{Deng et~al.(2025)Deng, Shen, Xie, Huang, Liang, and Wu}]{deng2025economic}
\bibinfo{author}{H.~Deng}, \bibinfo{author}{J.~Shen}, \bibinfo{author}{P.~Xie}, \bibinfo{author}{H.~Huang}, \bibinfo{author}{Z.~Liang}, \bibinfo{author}{H.~Wu},
\newblock \bibinfo{title}{Economic dispatch strategy for wind-containing power systems considering cascading failures},
\newblock \bibinfo{journal}{Electrical Engineering} \bibinfo{volume}{107} (\bibinfo{year}{2025}) \bibinfo{pages}{6123--6133}.
%Type = Article
\bibitem[{Liu et~al.(2022)Liu, Wang, Ju, and Li}]{liu2022sequentially}
\bibinfo{author}{H.~Liu}, \bibinfo{author}{C.~Wang}, \bibinfo{author}{P.~Ju}, \bibinfo{author}{H.~Li},
\newblock \bibinfo{title}{A sequentially preventive model enhancing power system resilience against extreme-weather-triggered failures},
\newblock \bibinfo{journal}{Renewable and Sustainable Energy Reviews} \bibinfo{volume}{156} (\bibinfo{year}{2022}) \bibinfo{pages}{111945}.
%Type = Article
\bibitem[{Ouyang et~al.(2023)Ouyang, Yu, Long, Diao, and Wang}]{10374403}
\bibinfo{author}{J.~Ouyang}, \bibinfo{author}{J.~Yu}, \bibinfo{author}{X.~Long}, \bibinfo{author}{Y.~Diao}, \bibinfo{author}{J.~Wang},
\newblock \bibinfo{title}{Coordination control method to block cascading failure of a renewable generation power system under line dynamic security},
\newblock \bibinfo{journal}{Protection and Control of Modern Power Systems} \bibinfo{volume}{8} (\bibinfo{year}{2023}) \bibinfo{pages}{1--11}.
%Type = Article
\bibitem[{Liang et~al.(2023)Liang, Guo, Zocca, Low, and Wierman}]{9872085}
\bibinfo{author}{C.~Liang}, \bibinfo{author}{L.~Guo}, \bibinfo{author}{A.~Zocca}, \bibinfo{author}{S.~Low}, \bibinfo{author}{A.~Wierman},
\newblock \bibinfo{title}{Adaptive network response to line failures in power systems},
\newblock \bibinfo{journal}{IEEE Transactions on Control of Network Systems} \bibinfo{volume}{10} (\bibinfo{year}{2023}) \bibinfo{pages}{333--344}.
%Type = Article
\bibitem[{Hu et~al.(2026)Hu, Zhao, Zhang, Xu, Chen, and Chen}]{11654520}
\bibinfo{author}{Z.~Hu}, \bibinfo{author}{Y.~Zhao}, \bibinfo{author}{W.~Zhang}, \bibinfo{author}{Z.~Xu}, \bibinfo{author}{Y.~Chen}, \bibinfo{author}{C.~Chen},
\newblock \bibinfo{title}{Proactive resilient uc against typhoon-induced cascading failures via a data-driven surrogate},
\newblock \bibinfo{journal}{IEEE Transactions on Power Systems}  (\bibinfo{year}{2026}) \bibinfo{pages}{1--4}.
%Type = Misc
\bibitem[{Donnot(2020)}]{grid2op}
\bibinfo{author}{B.~Donnot}, \bibinfo{title}{Grid2op: A testbed platform to model sequential decision making in power systems}, \bibinfo{year}{2020}. \URLprefix \url{https://github.com/Grid2op/grid2op}.

\end{thebibliography}

\end{document}